\documentclass[prb,twocolumn,aps,floats,floatfix,superscriptaddress]{revtex4}
\usepackage{epsfig}
\usepackage{colordvi}
\usepackage{graphicx}
\usepackage{amssymb}
\usepackage{color}

\begin{document}
\title{Plasmonically assisted channels of photoemission from metals}
\author{Dino Novko\footnote{Email: dnovko@ifs.hr }}
\affiliation{Institute of Physics, HR 10000 Zagreb, Croatia}
\author{Vito Despoja}
\affiliation{Institute of Physics, HR 10000 Zagreb, Croatia}
\author{Marcel Reutzel}
\affiliation{I. Physikalisches Institut, Georg-August-Universit\"{a}t G\"{o}ttingen, D-37077 G\"{o}ttingen, Germany}
\author{Andi Li}
\affiliation{Department of Physics and Astronomy and Pittsburgh Quantum Institute, University of Pittsburgh,
Pittsburgh, Pennsylvania 15260, USA}
\author{Hrvoje Petek}
\affiliation{Department of Physics and Astronomy and Pittsburgh Quantum Institute, University of Pittsburgh,
Pittsburgh, Pennsylvania 15260, USA}
\author{Branko Gumhalter\footnote{Corresponding author. Email: branko@ifs.hr}}
\affiliation{Institute of Physics, HR 10000 Zagreb, Croatia}

\begin{abstract}
We analyse recently measured nonlinear photoemission spectra from Ag surfaces that reveal resonances whose energies do not scale with the applied photon energy but stay pinned to multiples of bulk plasmon energy $\hbar\omega_p$ above the Fermi level. To elucidate these unexpected and peculiar features we investigate the spectra of plasmons generated in a solid by the optically pumped electronic polarization and their effect on photoemission. By combining quadratic response formalism for calculations of photoemission yield, a nonperturbative approach to inelastic electron scattering, and first-principles calculations for the electronic structure, we demonstrate the dependence of probability amplitude for single- and multiplasmon excitations on the basic parameters characterizing the photon pulse and the system. The resulting multiexcitation spectrum evolves towards a truncated plasmonic coherent state. Analogous concept is extrapolated to interpret plasmon generation by multiphoton excited electronic polarization. Based on this we elaborate a scenario that the thus created real plasmons act as supplementary frequency-locked pump field for non-Einsteinian plasmonically assisted channels of photoemission from metals. The established paradigm enables assignment and assessment of the observed linear $\hbar\omega_p$- and nonlinear $2\hbar\omega_p$-electron yields from Ag. Such effects may be exploited for selective filtering of optical energy conversion in electronic systems. 
\end{abstract}

\date{\today}
\maketitle

\newcommand{\bq}{\begin{equation}}
\newcommand{\eq}{\end{equation}}

\newcommand{\barr}{\begin{eqnarray}}
\newcommand{\earr}{\end{eqnarray}}

\section{Introduction}
\label{sec:introduction}

\begin{figure}[tb]
\rotatebox{0}{\epsfxsize=8 cm \epsffile{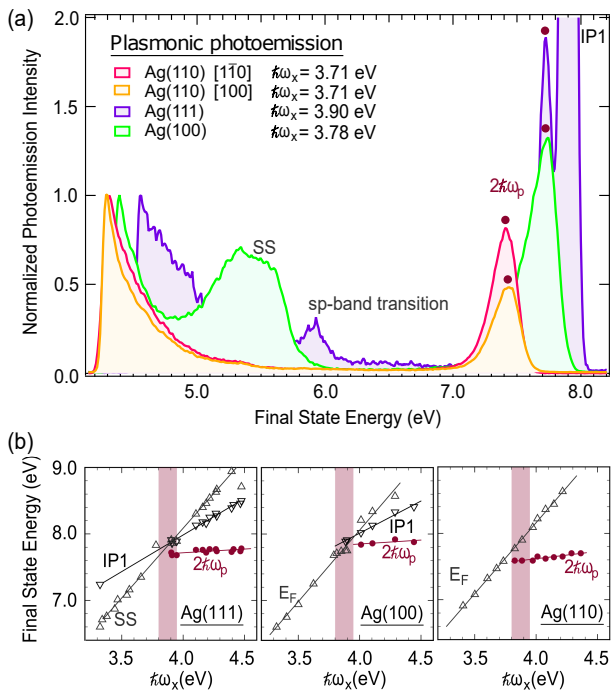}} 
\caption{Panel (a): Selection of 2PP spectra from three low index surfaces of Ag for excitation field energies $\hbar\omega_x$ in the ENZ region where the real part of Ag  dielectric function $\varepsilon(\omega)$ is near zero.[\onlinecite{ACSPhotonics}]  The manifolds of all the dots shown above the four representative plasmonic peak maxima make the $2\hbar\omega_p$ scaling curves in panel (b) below.
Panel (b): Dependence of the measured final state energies of 2PP electron yields from Ag on the variation of energy $\hbar\omega_x$. The data shown by triangles were recorded in the constant inital state mode (CIS) involving the surface state (SS) and the first image potential state (IP1) on Ag(111), the IP1 and the Fermi level $E_F$ on Ag(100), and $E_F$ on Ag(110). They exhibit the standard scaling of 2PP yield energy with $2\hbar\omega_x$ in emission from SS and $E_F$, and with $\hbar\omega_x$ in emission from IP1. In contrast to this, there also appear yields with energy $\sim 2\hbar\omega_p$ above $E_F$ (dots) which do not scale with the multiples of the radiation field energy $p\hbar\omega_x$. This $2\hbar\omega_p$ emission extends far beyond the ENZ region denoted by vertical shading.}
\label{AllAg}
\end{figure}

In the standard one-electron picture of linear photoeffect the kinetic energy of electrons photoemitted from the system scales with the energy of the quanta of applied electromagnetic (EM) field following Einstein's relation
\bq
\epsilon_f=\hbar\omega_x+\epsilon_B.
\label{eq:Einstein}
\eq
Here $\epsilon_f$ and $\epsilon_B$ are the final kinetic and the initial binding energy of the electron, respectively,  and $\hbar\omega_x$ is the photon energy. In nonlinear multiphoton photoemission ($m$PP, $m\geq 2$)  that measures the dynamics of photoexcitation and unoccupied electronic structure of solids\cite{ReutzelPRX}
$\epsilon_f$ scales with  multiples $p\hbar\omega_x$ ($1\leq p \leq m$) of the photon energy, with $p$ depending on (i.e. decreasing with) the number of resonant intermediate states partaking in the transition (see Fig.\,2 in Ref. [\onlinecite{TM}]).  Novel measurements of two-photon photoemission (2PP) spectra from Ag(111), Ag(100) and Ag(110) surfaces\cite{MarcelPRL,AndiNJP,ACSPhotonics} have confirmed with high resolution the obeyance of such scalings in one-electron transitions which broadly determine the 2PP spectra from metallic band states.\cite{Adawi,FausterSteinmann,PetekOgawa,Weinelt,TimmBennemann,UebaGumhalter} However, besides this, they have also revealed peaks in photoemission yield at electron kinetic energies which do not scale with the photon energy of applied EM field as would be expected from the extension of Einsteinian relation (\ref{eq:Einstein}) to 2PP. Instead, these peaks occur as resonances pinned to the double bulk plasmon energy $2\hbar\omega_p$ above the Fermi level $E_F$ and their energy appears independent of the energy of excitation source as long as $\hbar\omega_x\geq \hbar\omega_p\sim 3.8$ eV. This peculiar behaviour is illustrated in Fig.\,\ref{AllAg}. It can not be interpreted as a standard $n$-th plasmon satellite whose energy would exhibit a combined ($p,n$) scaling:
\bq
\epsilon_{sat}=p\hbar\omega_{x}+\epsilon_{B}-n\hbar\omega_{p}.
\label{eq:satellite}
\eq
 Identical finding, although differently interpreted,  was reported at the dawn of 2PP spectroscopy for the Ag(111) surface.\cite{Steinmann85}  Analogous feature is also implicit in  one-photon photoemission (1PP) spectra from submonolayer sodium coated Ag(100) recorded in the constant initial state (CIS) mode which exhibit peaks pinned to $\hbar\omega_p$ above $E_F$.\cite{RefHorn,Horn} These intriguing "non-Einsteinian" photoemission modes call for interpretations going beyond the sole effect of electron excitation by dielectrically screened EM fields\cite{Feibelman,Forstmann,Krasovskii,Reshef} which oscillate with the frequencies of the applied external fields. Evidently, the novel paradigm should involve real plasmon excitation and energy exchange in the intermediate and final stages of one-electron emission processes. The nonthermally excited plasmons whose energy largely exceeds the thermal energy $kT$ may like electrons\cite{encyclopedia} also be termed "hot". Hence a proof of concept is needed to ascertain as how the primary optically induced electronic polarization may generate hot plasmons that via subsequent absorption can supply energy for non-Einsteinian electron emission from the irradiated system. 

The possibility that real plasmons appear as intermediaries in photoemission and photoabsorption was alluded to long ago.\cite{Hopfield1965,Tzoar,Gornyi,Kliewer} 
More recently, and in the context of ultrashort pulse photoexcitations, it was noted in Ref. [\onlinecite{Schueler}]. Expanding on this we first elaborate a microscopic mechanism of generation of real plasmons by light induced electronic polarization, i.e. by the primary optically excited electron-hole pairs.  This mechanism  is  complementary to direct plasmon excitation by local EM fields in microscopically inhomogeneous media.\cite{Feibelman,Forstmann,Podbiel} Once being excited in the system, real plasmons may support additional excitation pathways for the electrons that could sustain preferential narow band optical energy conversion.\cite{Linic,Boriskina,Narang2016} Such consecutive processes may become effective in linear and nonlinear photoemission by supplementing the primary hot electron excitation channels,\cite{Marini2002,Atwater14,Nordlander15,Louie15,Halas15,deAbajo16,Atwater16,Khurgin,Govorov17} and thereby modify the distributions of photoemission yields from metals.\cite{MarcelPRL,ACSPhotonics,Horn} To demonstrate this we first define a calculable measure of the spectral distribution of hot plasmons excitable in interactions of the electron system with external EM fields. Establishment of a distribution of real plasmons in the system enables us to exploit the analogy with $m$PP theory\cite{UebaGumhalter} and formulate the plasmon generated electron yields that supplement the standard $m$PP ones.

In Sec.\,\ref{sec:model} we introduce a minimal model for description of electron gas in interaction with plasmons and pulsed EM field, viz. with only two members of the plethora of excitation modes in polarizable media.\cite{Rivera} Using this model we demonstrate in Sec. \ref{sec:spectrum} a construction of quantum mechanical expression describing the spectrum of real plasmons generated in the system by optically induced electronic polarization.  The construction is explicitly shown for 1PP in which case it combines the quadratic response to EM field\cite{Ashcroft,Mahan1PP,Caroli} with the nonperturbative scattering formalism for description of multiboson excitation processes.\cite{PhysRep} They are respectively used to formulate the preparation of electronic polarization excited by EM field and subsequent polarization-induced generation of plasmonic coherent states over the electronic state of the system. Hot electrons arising from spontaneous decay of excited plasmons complement the preferential photoelectron yield from the Fermi level\cite{Hopfield1965} at the multiples of plasmon energy and thereby explain the non-Einsteinian features of $m$PP spectra reported in Refs. [\onlinecite{MarcelPRL,ACSPhotonics,Steinmann85,Horn}]. 
Special limits of the obtained results that are of general interest are explored and plasmon generation probabilities  derived and discussed in Sec. \ref{sec:application}. Expanding on this we propose in Sec. \ref{sec:plasmonpump} a theoretical framework in which  plasmons excited in multiphoton driven polarization processes may give rise to plasmon frequency-locked electron emission from irradiated solids. In Sec. \ref{sec:results} we employ the density functional theory to calculate the relevant rates of consecutive plasmon and hot electron generation in Ag bands that illustrate the scenarios for non-Einsteinian channels of 1PP and 2PP from metals. This provides essential prerequisites for the studies of plasmonically assisted channeling of broad band optically induced electronic excitations into a narrow energy interval spanned only by the range of Ag plasmon dispersion.

\section{Excitation of plasmons by optically induced electronic polarization}
\label{sec:System} 
\subsection{Description of the model}
\label{sec:model}

We start with a quantum description of light-matter interaction\cite{josab} based on a simple model Hamiltonian $H$ comprising the component that describes the unperturbed system, $H^{syst}$, and coupling $W(t)$ of the system to EM field 
\bq
H=H^{syst}+W(t).
\label{eq:H}
\eq
Here $H^{syst}$ is composed of $H_{0}^{e}$ describing electrons in crystal band(s), $H_{0}^{pl}$ describing unperturbed plasmons, and $V^{e-pl}$ describing their interaction, viz.
\bq
H^{syst}=H_{0}^{e}+H_{0}^{pl}+V^{e-pl}=H_{0}^{syst}+V^{e-pl}.
\label{eq:Hsyst}
\eq
In the second quantization they are respectively given by 

\barr
H_{0}^{e} &=& \sum_{\bf k}\epsilon_{\bf k}c_{\bf k}^{\dag}c_{\bf k}, 
\label{eq:allHel}\\
H_{0}^{pl} &=& \sum_{\bf q}\hbar\omega_{\bf q}a_{\bf q}^{\dag}a_{\bf q}, 
\label{eq:allHpl}\\
 V^{e-pl} &=& g\sum_{\bf k,\bar{k},q}V_{\bf \bar{k},k}^{\bf q}c_{\bf \bar{k}}^{\dag}c_{\bf k}a_{\bf q}^{\dag} +h.c., 
\label{eq:allHint}
\earr
where  $c_{\bf k}^{\dag}$ and $c_{\bf k}$ ($a_{\bf q}^{\dag}$ and $a_{\bf q}$) denote the electron (plasmon) creation and destruction operators, respectively, which obey the fermion (boson) commutation rules. $\epsilon_{\bf k}$ ($\hbar\omega_{\bf q}$) stand for the electron (plasmon) energy associated with the quantum numbers ${\bf k}$ (${\bf q})$ which may include the Bloch momentum, band index, etc., describing the unperturbed dynamics of quasiparticles in the system.  In this general notation $V_{\bf \bar{k},k}^{\bf q}$ is the electron-plasmon coupling matrix element and $g$ is the coupling constant eventually to be set equal to unity. For later convenience we assume that this interaction is switched on and off adiabatically, viz. $V^{e-pl}\propto e^{-\eta|t|}$ where $\eta\rightarrow 0^{+}$.

The electron coupling to the EM field is described by
\barr
W(t)&=&\sum_{i}{\cal E}_{i}(\omega_{x,i},t,\sigma_{x,i})e^{-i(\omega_{x,i}t+\varphi_{i})}
\nonumber\\
&\times&\sum_{\bf k_{1},k_{2}}W_{\bf k_{1},k_{2}}c_{\bf k_{1}}^{\dag}c_{\bf k_{2}}+h.c. 
\label{eq:W}
\earr
Here ${\cal E}_{i}(\omega_{x,i},t,\sigma_{x,i})>0$, $\omega_{x,i}>0$ and $\varphi_{i}$ are the amplitude modulation of the $i$-th pulse constituting the EM field (pulse profile centered at $t_{x,i}$ and of duration $\sigma_{x,i}$) and the pulse carrier frequency and phase in the crystal, respectively, and $W_{\bf k_1,k_2}$ denote the dipole matrix elements of the electron interaction with the field.

Quite generally, the field amplitudes ${\cal E}_{i}(\omega_{x,i},t,\sigma_{x,i})$ and the matrix elements $W_{\bf k_1,k_2}$ depend on the dielectric properties and the symmetry of the system that affect selection rules for ${\bf k_2\leftrightarrow k_1}$ field-induced electronic transitions.  
 Thus, the effective ${\cal E}_{i}(\omega_{x,i},t,\sigma_{x,i})$ would incorporate local and nonlocal Fresnel or Feibelman\cite{Feibelman} corrections for the EM field in the sample. It may also be strongly enhanced for metals at epsilon-near-zero\cite{Reshef} (ENZ) where the real component of linear  permitivity is zero. Effective $W_{\bf k_1,k_2}$ may also embody the surface enhanced\cite{Walden} and Drude-assisted\cite{KliewerBennemann} transitions in a phenomenological fashion. 
In the following we shall restrict our discussions to the quadratic and quartic response of the electronic system to action of one pulse of the EM field. In this $i=1$ limit of (\ref{eq:W}) the formal derivations of the various 1PP- and 2PP-related transition rates can be demonstrated in a conscpicuous fashion. Nonlinear responses induced by two pulses ($i=1,2$) and leading to 2PP and 3PP photoemission yields that are unaffected by real plasmon excitations were studied numerically in the Supplementary Material of Ref. [\onlinecite{TM}]. 

\subsection{Formulation of the spectrum of hot plasmons generated by photoexcited quasiparticles}
\label{sec:spectrum}

We seek a scenario based on the minimal electron-plasmon model outlined in Sec. \ref{sec:model} that may enable interpretation of the plasmon energy locked photoelectron yields for  $\omega_x\geq \omega_p$ reported previously\cite{MarcelPRL,ACSPhotonics,Steinmann85,Horn} and here in Fig. \ref{AllAg} for three low index Ag surfaces. Such yields occuring at multiples $n\hbar\omega_p$ of plasmon energy indicate the pumping of electrons by hot plasmons generated in the course of photoemission process. An example of lowest order process of this kind ($n=1$), which starts from the electron-plasmon system in the ground state that is linearly perturbed by the EM field (\ref{eq:W}), can be symbolically represented by the flow scheme

\begin{figure}[tb]
\rotatebox{0}{\epsfxsize=7.5cm \epsffile{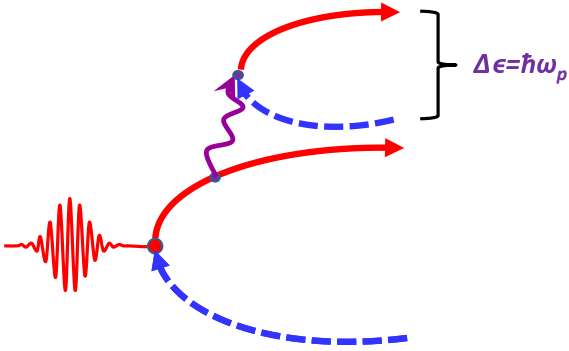}} 
\caption{Diagrammatic representation of the sequence of excitation events (\ref{eq:sequence}) linearly induced in the electron-plasmon system by pulsed EM field. Full red, dashed blue and purple wiggly lines denote electron, hole and plasmon propagation, respectively. Blue and red circles denote matrix elements $V_{\bf \bar{k_{1}},k_{1}}$ and $W_{\bf k_{1},k_{2}}$ defined in Eqs.  (\ref{eq:allHint}) and (\ref{eq:W}), respectively. The difference of quasiparticle energies of the secondary electron and hole arising from real plasmon absorption is $\Delta\epsilon \simeq \hbar\omega_p$. }
\label{sequence}
\end{figure}

\begin{widetext}
\barr
& &n=1:
\nonumber\\
& &|\mbox{electronic grd. state}\rangle|\rm{photon}\rangle
{\stackrel{\it W}{\scriptscriptstyle \longrightarrow}}
|\rm{polarization}\rangle
{\stackrel{\it V^{e-pl}}{\scriptscriptstyle \longrightarrow}}
|\rm{polarization'}\rangle|\rm{plasmon}_{\hbar\omega_p}\rangle
{\stackrel{\it V^{pl-e}}{\scriptscriptstyle \longrightarrow}}
|\rm{polarization''}\rangle|\rm{electron_{\hbar\omega_p}}\rangle
\label{eq:sequence}
\earr
\end{widetext}
with symbols over arrows denoting the interactions in (\ref{eq:H}) that drive particular transitions. Here $|\rm{polarization}\rangle
$, $|\rm{polarization'}\rangle$ and $|\rm{polarization''}\rangle$ denote the excited states of the electronic system with (i) one primary photoexcited electron above and a primary hole below $E_F$, (ii) primary electron scattered to a secondary state by emission of a real plasmon and the unperturbed primary hole, and (iii) ionized state of the system with secondary electron, primary hole and secondary hole created by plasmon absorption that gives rise to electron excitation to an emanating state $|{\rm electron_{\hbar\omega_p}}\rangle$ whose energy exceeds the initial one by $\hbar\omega_p$. This propagation sequence can be also illustrated by open diagram\cite{DuBois} shown in Fig. \ref{sequence}. For secondary holes created at the Fermi surface, the energy of the secondary emanating electrons is $\hbar\omega_p$-locked above $E_F$. Analogous sequence can be established for the case of plasmon generation by primary hole scattering.

As is evident from the sequence (\ref{eq:sequence}) and the ensuing description, the key intermediate events determining the final $\hbar\omega_p$-locked electron yield are the  emissions of real plasmons which likewise photons can convey their energy to excite electrons emanating from the system. Therefore, we need as a prerequisite a rigorous assesment of multiple excitation of real plasmons using the above framework for light-matter interaction. To this end we define the multiplasmon spectrum ${\cal S}^{pl}(\varepsilon,t)$ as composed of energies $\varepsilon$ of all real plasmons generated during the propagation of electronic polarization excited by the interaction of electrons with the pulsed EM field (\ref{eq:W}). This is obtained from the expresssion
\bq
{\cal S}^{pl}(\varepsilon,t)=\langle\Psi(t)|\delta(\varepsilon - H_{0}^{pl})|\Psi(t)\rangle, 
\label{eq:N}
\eq
in which the function $\delta(\varepsilon - H_{0}^{pl})$ acts as a "plasmon counter" operator that projects from the wavefunction $|\Psi(t)\rangle$ of the laser-perturbed system all the states with multiply excited bosonic energies $n\hbar\omega_{\bf q}$, each component being weighted by the corresponding boson excitation probability. This expression for plasmon counting has an advantage over the form 
\bq
{\cal N}^{pl}(N,t)=\langle\Psi(t)|\delta(N - \sum_{\bf q} a_{\bf q}^{\dag}a_{\bf q})|\Psi(t)\rangle, 
\label{eq:Ndirect}
\eq
which directly gives the number $N=\sum_{\bf q}n_{\bf q}$ of excited bosons. The latter may diverge in the case of linear boson dispersion and transient and localized perturbations.\cite{Hopfield} By contrast, expression (\ref{eq:N}) renders regular results also for linearly dispersing bosons.  
%

\begin{figure}[tb]
\rotatebox{0}{\epsfxsize=8 cm\epsffile{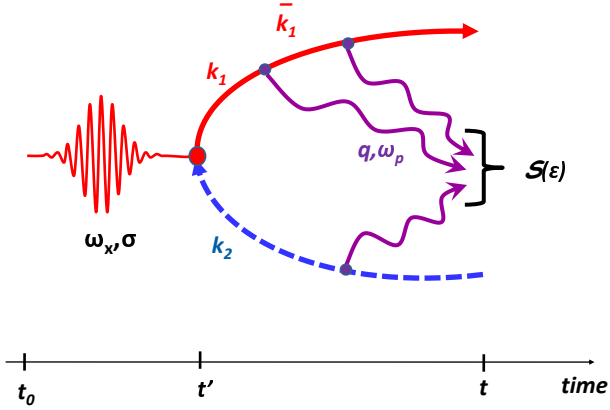}} 
\caption{Diagrammatic representation of linearly induced plasmon generation processes described by (\ref{eq:N}) with $|\Psi(t)\rangle$ defined in  (\ref{eq:Psi}) and leading to (\ref{eq:Nplkk}). Meaning of the symbols is same as in Fig. \ref{sequence}. The time evolution after the application of pulsed perturbation  (\ref{eq:W}) is governed by the Hamiltonian (\ref{eq:Hsyst}).  Plasmon counting implemented through the operator $\delta(\varepsilon-H_{0}^{pl})$ is denoted symbolically by ${\cal S}(\varepsilon)$. Real multiplasmon excitations may occur for excited quasiparticle energies exceeding the multiexcitation threshold $n\omega_p$.}
\label{Psi}
\end{figure}

\begin{figure}[tb]
\rotatebox{0}{\epsfxsize=6cm\epsffile{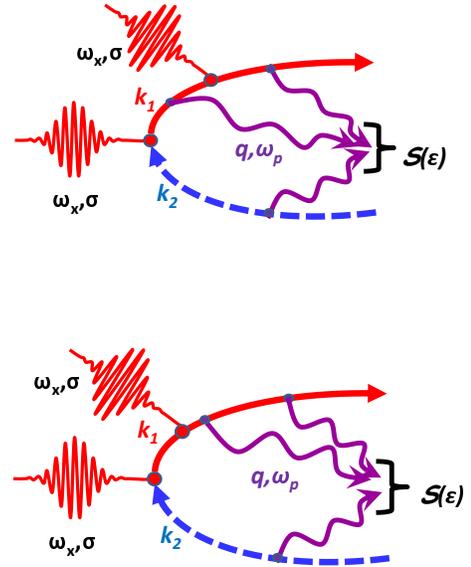}} 
\caption{Diagrammatic representation of multiplasmon generation driven by electronic polarizarion excited by double action of the laser pulse. All symbols have the same meaning as in Fig. \ref{Psi}. Electron induced plasmon generation  may take place either in the intermediate and final electron states (upper panel), or in the final state of the two photon-driven process (lower panel).}
\label{Psi2P}
\end{figure}

To derive an explicit form of (\ref{eq:N}) we need to find a convenient and realistic representation of the excited state wavefunction $|\Psi(t)\rangle$ and to facilitate this we introduce several simplifying assumptions. The preparation of  excited state starts from the initial unperturbed ground state of the system $|0\rangle$ at some instant $t_{0}$ in the remote past before the perturbation $W(t)$ has been switched on. Real plasmon generation driven by one photon- and two photon-induced polarization are illustrated with open diagrams\cite{DuBois} in Figs. \ref{Psi} and \ref{Psi2P}. However, since the "non-Einsteinian peaks" have been observed already in 1PP spectra,\cite{Horn} in order to present a proof of concept for their occurence it suffices to employ the general framework of linear photoemission theory\cite{Ashcroft} and calculate (\ref{eq:N}) in the quadratic response with respect to perturbation (\ref{eq:W}). This amounts to acting at the time $t'<t$ with $W(t')$ on the  state $U^{syst}(t',t_{0})|0\rangle$ only once (one photon absorption driven processes) and then letting the thus excited system to evolve under the action of the Hamiltonian (\ref{eq:Hsyst}). This interaction sequence is symbolically represented by the diagram in Fig. \ref{Psi}.
This produces $|\Psi(t)\rangle$ in the form
\bq
|\Psi(t)\rangle= \int_{t_0}^{t}\frac{dt'}{i\hbar}U^{syst}(t,t')W(t')U^{syst}(t',t_{0})|0\rangle, 
\label{eq:Psifull}
\eq
where  $U^{syst}(t,t')$ is the evolution operator in the Schr\"{o}dinger picture of the system in the absence of interaction $W(t')$, viz.
\bq
U^{syst}(t,t')= \exp[-\frac{i}{\hbar}H^{syst}(t-t')]. 
\label{eq:Usyst}
\eq
Note in passing that in accord with Fig. \ref{Psi} expression (\ref{eq:Psifull}) defines the wave function of an open system describing the processes of photon absorption and subsequent real plasmon generation 
and not the steady state evolution of electron-plasmon system discussed in Ref. [\onlinecite{PS}].

Expression (\ref{eq:N}) with substitution (\ref{eq:Psifull}) is exact within the quadratic response but still operationally too complicated. Hence, at this stage we shall invoke the next simplifying assumption. As we are interested only in real plasmon generation we shall neglect the fluctuations of the ground state (i.e. the Fermi sea) induced by electron-plasmon interaction $V^{e-pl}$ in $U^{syst}(t',t_{0})$ on the RHS of the integrand in (\ref{eq:Psifull}), and replace it by 
\bq
U_{0}^{syst}(t,t')= \exp[-\frac{i}{\hbar}H_{0}^{syst}(t-t')],
\label{eq:U0syst}
\eq
where $H_{0}^{syst}=H_{0}^{e}+H_{0}^{pl}$. This yields a simpler and operationally more convenient expression 
\bq
|\Psi(t)\rangle= \int_{t_0}^{t} \frac{dt'}{i\hbar}U^{syst}(t,t')W(t')U_{0}^{syst}(t',t_{0})|0\rangle, 
\label{eq:Psi}
\eq
in which $U_{0}^{syst}(t',t_{0})$ is diagonal on $|0\rangle$, viz. $U_{0}^{syst}(t',t_{0})|0\rangle=e^{-\frac{i}{\hbar}E_{0}(t'-t_{0})}|0\rangle$ where $E_{0}$ is the ground state energy, hereafter set equal to zero.

Introducing the Fourier transform 
\bq
\delta(\varepsilon - H_{0}^{pl})=(2\pi\hbar)^{-1}\int_{-\infty}^{\infty}d\tau e^{\frac{i}{\hbar}(\varepsilon -H_{0}^{pl})\tau}
\label{eq:delta}
\eq
we can bring (\ref{eq:N}) after suppressing index $i$ in (\ref{eq:W}) to the form

\begin{widetext}
\barr
{\cal S}^{pl}(\varepsilon,t)&=&\frac{1}{2\pi\hbar^3}\int_{-\infty}^{\infty}d\tau e^{i\varepsilon\tau}\int_{t_0}^{t} dt''\int_{t_0}^{t} dt'\langle 0|W(t'')U^{syst}(t,t'')^{\dag}e^{-\frac{i}{\hbar}H_{0}^{pl}\tau}U^{syst}(t,t')W(t')|0\rangle 
\label{eq:Npl}\\
&=&
\frac{1}{2\pi\hbar^3}\int_{-\infty}^{\infty}d\tau e^{i\varepsilon\tau}\int_{t_0}^{t}{\cal E}(\omega_x,t'',\sigma_{x})e^{i\omega_{x}t''} dt''\int_{t_0}^{t}{\cal E}(\omega_x,t',\sigma_{x})e^{-i\omega_{x}t'} dt'\nonumber\\
&\times&
\sum_{\bf k_{1}'',k_{2}'',k_{1}',k_{2}'} W_{\bf k_{1}'',k_{2}''}W_{\bf k_{2}',k_{1}'}\langle 0|c_{\bf k_{2}''}^{\dag}c_{\bf k_{1}''}U^{syst}(t,t'')^{\dag}e^{-\frac{i}{\hbar}H_{0}^{pl}\tau}U^{syst}(t,t')c_{\bf k_{1}'}^{\dag}c_{\bf k_{2}'}|0\rangle.
\label{eq:Nplkk} 
\earr
\end{widetext}
Note here that $\tau$ appearing in $e^{-\frac{i}{\hbar}H_{0}^{pl}\tau}$ is not associated with any time evolution of the system but only plays the role of integration variable conjugate to the projected energy $\varepsilon$, whereas the real evolution times are $t,t',t''$.

The main complications in evaluating (\ref{eq:Nplkk}) arise with the treatment of matrix  elements of the products of three generalized exponential operators in the second line of (\ref{eq:Nplkk}). This is addressed in the next subsection.

\subsection{Approximations and operator algebra in the calculations of hot plasmon spectrum in  Eq. (\ref{eq:Nplkk})}
\label{sec:approximations}

We seek a nonperturbative solution in terms of $V^{e-pl}$ to expression in the second line on the RHS of (\ref{eq:Nplkk}). Here the sum $\sum_{\bf k_{1}'',k_{2}'',k_{1}',k_{2}'}...$ comprises coherent (${\bf k_{1'}=k_{1''}}$ and ${\bf k_{2'}=k_{2''}}$) and incoherent components (${\bf k_{1'}\neq k_{1''}}$ and ${\bf k_{2'}\neq k_{2''}}$) arising in the calculation of the ground state average in (\ref{eq:Nplkk}). The evaluation of all such components represents a formidable task but simplifications which enable grasping the most salient features of hot plasmon generation are possible. For this sake we shall  restrict the  plasmon generation to processes in which inelastic electron scattering above the Fermi level  is dominant (i.e. $\epsilon_{\bf k}>E_{F}, \epsilon_{\bf \bar{k}}>E_{F}$ after the action of $V^{e-pl}$ on the photoexcited state), and  neglect plasmon generation by holes excited below $E_{F}$.  This is justified for the CIS mode photoyield measured from the vicinity of the Fermi surface (cf. Fig. 7 of Ref. [\onlinecite{Horn}]) in which case the processes involving the hole recoil (i.e. ${\bf k_{2}'\neq k_{2}''}$) due to real plasmon emission must be excluded. This is also consistent with the earlier exclusion of  Fermi sea fluctuations in the interval $(t_0,t')$ in Eq. (\ref{eq:Psi}). A quantitative support for such approximation will be provided in Sec. \ref{sec:results} by the calculations of plasmon generation by electrons and holes propagating in Ag bands.

By restricting the plasmon coupling to only one type of quasiparticle we omit all plasmon mediated excitonic effects associated with the propagation of  primary photoexcited e-h pair\cite{Mahan,Gavoret,Combescot,Marini,PSS} but this restriction does not invalidate the desired proof of concept of hot plasmon generation in solids. In the opposite limit of dominant plasmon-hole coupling a completely analogous procedure of neglecting electron-induced plasmon generation can be followed.\cite{Reining,FCaruso,GW+C,Reining2018} Moreover, for $\omega_p \leq\omega_x<2\omega_p$ (experimental conditions in Fig. \ref{AllAg}) only one of these two hot plasmon generation channels can be effective.

In the next step, and in accord with Ref. [\onlinecite{Ashcroft}], we only consider such coherent contributions from (\ref{eq:Nplkk}) which give rise to steady state quantities in the long time limit, here effective already past the duration of the pulse. This renders ${\bf k_{1}''=k_{1}'=k_{1}}$, ${\bf k_{2}''=k_{2}'=k_{2}}$ and reduces the calculation of (\ref{eq:Nplkk}) to a much simpler quantity in which, as regards the interaction with plasmons, the electronic states above and below $E_F$ are decoupled except for the relation between ${\bf k_{1}}$ and ${\bf k_{2}}$ imposed by the selection rules contained in the optical transition matrix elements $W_{\bf k_{1},k_{2}}$. In this case the ground state average in (\ref{eq:Nplkk}) factorizes into diagonal electron and hole components
%

\barr
& &\langle 0|c_{\bf k_{2}}^{\dag}c_{\bf k_{1}}U^{syst}(t,t'')^{\dag}e^{-\frac{i}{\hbar}H_{0}^{pl}\tau}U^{syst}(t,t')c_{\bf k_{1}}^{\dag}c_{\bf k_{2}}|0\rangle\nonumber\\
&=&
\langle 0|c_{\bf k_{2}}^{\dag}(t'')c_{\bf k_{2}}(t')|0\rangle\nonumber\\
&\times&
\langle 0|c_{\bf k_{1}}U^{syst}(t,t'')^{\dag}e^{-\frac{i}{\hbar}H_{0}^{pl}\tau}U^{syst}(t,t')c_{\bf k_{1}}^{\dag}|0\rangle.
\label{eq:Ndiag} 
\earr
%
Here $\langle 0|c_{\bf k_{2}}^{\dag}(t'')c_{\bf k_{2}}(t')|0\rangle=n_{\bf k_{2}}\exp[i\epsilon_{\bf k_{2}}(t''-t')]$ is the unperturbed hole propagator or correlation function with $n_{\bf k_{2}}=\langle 0|c_{\bf k_{2}}^{\dag}c_{\bf k_{2}}|0\rangle$, and the electron component appears in the form of an excited state average of a product of generalized exponential operators. The coupled photoexcited electron-plasmon entity is usually termed plasmonic polaron or plasmaron.\cite{Tediosi,Chis}  

The electron component on the RHS of (\ref{eq:Ndiag}),viz. 
\barr
& &\langle 0|c_{\bf k_{1}}U^{syst}(t'',t)^{\dag}e^{-\frac{i}{\hbar}H_{0}^{pl}\tau}U^{syst}(t,t')c_{\bf k_{1}}^{\dag}|0\rangle\nonumber\\
&=&
\langle{\bf k_{1}}|U^{syst}(t'',t)^{\dag}e^{-\frac{i}{\hbar}H_{0}^{pl}\tau}U^{syst}(t,t')|{\bf k_{1}}\rangle 
\label{eq:electronpart}
\earr
acquires the form of the counting-intercepted (via the action of $e^{-\frac{i}{\hbar}H_{0}^{pl}\tau}$) plasmaron propagator which describes spontaneous emissions of plasmons\cite{PS} by the primary electron photoexcited to the state $|{\bf k_{1}}\rangle$. Its evaluation proceeds by formal manipulations of the three generalized exponential operators partaking in this excited state average. The details of corresponding calculations are presented in Appendix \ref{sec:ElectronPart}. 
Making use of expression (\ref{eq:<eWfinal>}) derived therein we obtain the leading contribution to  (\ref{eq:Nplkk}) that is quadratic (i.e. perturbative) in the EM field, while the effect of electron-plasmon coupling $V_{\bf \bar{k}_{1},k_{1}}^{\bf q}$ is formally included to all orders (i.e. nonperturbatively).  
Introducing the transient integral 
\bq
\Delta_{\bf \bar{k},k}^{\bf q}(t,t')=\frac{1}{\hbar}
\int_{t'}^{t}e^{-\frac{i}{\hbar}(\epsilon_{\bf k}-\epsilon_{\bf \bar{k}}-\hbar\omega_{\bf q})t_{1}}dt_{1}
\label{eq:DeltaText}
\eq
the afore mentioned leading contribution reads

\begin{widetext}
\barr
{\cal S}^{pl}(\varepsilon,t)
&=&
\frac{1}{\hbar^2}\int_{t_0}^{t}{\cal E}(\omega_x,t'',\sigma_{x}) dt''\int_{t_0}^{t}{\cal E}(\omega_x,t',\sigma_{x}) dt'
\sum_{\bf k_{1},k_{2}} |W_{\bf k_{1},k_{2}}|^{2}(1-n_{\bf k_{1}})n_{\bf k_{2}}e^{\frac{i}{\hbar}(\hbar\omega_{x} -\epsilon_{\bf k_{1}}+\epsilon_{\bf k_{2}})(t''-t')}\nonumber\\
&\times&
\frac{1}{2\pi\hbar}\int_{-\infty}^{\infty}d\tau e^{\frac{i}{\hbar}\varepsilon\tau}\exp\left[-\sum_{\bf \bar{k}_1,q}\left|V_{\bf \bar{k}_1,k_{1}}^{\bf q}\Delta_{\bf k_{1},\bar{k}_1}^{\bf q}(t,(t'+t'')/2)\right|^{2}(1-n_{\bf \bar{k}_1})(1-e^{-i\omega_{\bf q}\tau})\right],
\label{eq:Nplexp} 
\earr
\end{widetext}
with a proviso $\epsilon_{\bf k_{1}},\epsilon_{\bf \bar{k}_1}>E_{F}$. This and the symmetry selection rules contained in $V_{\bf \bar{k}_1,k_1}^{\bf q}$ restrict the ${\bf q}$-summation in the last line on the RHS of (\ref{eq:Nplexp}).

The quadratic response formalism here applied to description of electron photoexcitation temporally restricted to the pulse duration,  i.e. short times around $t_{x}$, required the introduction of finite times  $t$, $t'$ and $t''$ in the general expression (\ref{eq:Nplkk}).
Such dependence on the intermediate propagation times obviously renders  the spectra (\ref{eq:Nplexp}) of plasmons  emitted by photoexcited electrons much more complicated than would have been obtained within the standard scattering formalism.\cite{PhysRep,GW} The same formulation can be extended to quartic response appropriate to 2PP\cite{UebaGumhalter,encyclopedia} but this is technically, although not so much conceptually, more demanding.

 Lastly, we observe that in the absence of electron-plasmon coupling in (\ref{eq:Hsyst}) one obtains $w_{\bf k_{1}}(t,t',t'')=0$, the last line on the RHS of (\ref{eq:Nplexp}) is replaced by $\delta(\varepsilon)$ and the remainder yields the transient electronic polarization spectrum shown in Fig. 3 of Ref. [\onlinecite{PSS}].

\section{Applications of the obtained results in special limits}
\label{sec:application}

In this section we illustrate the behaviour of expression (\ref{eq:Nplexp}) in simple limits of physical relevance that all demonstrate the formation of plasmonic coherent states\cite{Glauber} in electronic systems perturbed by ultrashort laser pulses.

\subsection{Illustration of generation of plasmonic coherent states in the limit of nondispersive plasmons }
\label{sec:nondispersive}

The multiplasmon features of (\ref{eq:Nplkk}) and (\ref{eq:Nplexp}) obtained from the  plasmon counting formula (\ref{eq:N}) can now be best illustrated by assuming nondispersive plasmon frequency $\omega_{\bf q}=\omega_{p}$ in the second line on the RHS of (\ref{eq:Nplexp}). This allows the ${\bf q}$-summations to be readily carried out to give 

\barr
w_{\bf k_{1}}(t,t',t'')=\sum_{\bf \bar{k}_1,q}\left|V_{\bf \bar{k}_1,k_{1}}^{\bf q}\Delta_{\bf \bar{k}_1,k_1}^{\bf q}(t,t'+t'')/2)\right|^{2}(1-n_{\bf \bar{k}_1})\nonumber\\
\rightarrow
\sum_{\bf \bar{k}_1}(1-n_{\bf \bar{k}_1}){\cal V}_{\bf \bar{k}_{1},k_{1}}^{2}\left|\int_{\frac{(t'+t'')}{2}}^{t}e^{-\frac{i}{\hbar}(\epsilon_{\bf k_{1}}-\epsilon_{\bf \bar{k}_1}-\hbar\omega_{p})t_{1}}\frac{dt_{1}}{\hbar}\right|^{2}.
\label{eq:w}
\earr
where ${\cal V}_{\bf \bar{k}_{1},k_{1}}^{2}=\sum_{\bf q}\left|V_{\bf \bar{k}_1,k_{1}}^{\bf q}\right|^2$.
Using this, expanding the exponential function in the last line on the RHS of (\ref{eq:Nplexp}) into a power series and taking its Fourier transform term by term yields (with $t_{0}\rightarrow -\infty$)

\begin{widetext}
\barr
{\cal S}^{pl}(\varepsilon,t)&=&
\frac{1}{\hbar^2}\int_{-\infty}^{t}{\cal E}(\omega_x,t'',\sigma_{x}) dt''\int_{-\infty}^{t}{\cal E}(\omega_x,t',\sigma_{x}) dt'
\sum_{\bf k_{1},k_{2}} |W_{\bf k_{1},k_{2}}|^{2}(1-n_{\bf k_{1}})n_{\bf k_{2}}e^{-\frac{i}{\hbar}(\epsilon_{\bf k_{1}}-\epsilon_{\bf k_{2}}-\omega_{x})(t''-t')}\nonumber\\
&\times&
\sum_{n=0}^{\infty}e^{-w_{\bf k_{1}}(t,t',t'')}\frac{\left[w_{\bf k_{1}}(t,t',t'')\right]^{n}}{n!}\delta(\varepsilon-n\omega_{p}). 
\label{eq:Nplexpanded}
\earr
\end{widetext}
The last line on the RHS of (\ref{eq:Nplexpanded}) clearly demonstrates the signature of plasmon counting effectuated in ${\cal S}^{pl}(\omega,t)$ by the operator (\ref{eq:delta}) and manifesting through the integer $n$ ($0\leq n\leq \infty$) in the expansion of the diabatically developing time dependent plasmonic coherent state  

\bq
|w_{\bf k_{1}}(t,t',t'')\rangle=\sum_{n=0}^{\infty}e^{-w_{\bf k_{1}}(t,t',t'')/2}\sqrt{\frac{w_{\bf k_{1}}(t,t',t'')^{n}}{n!}}|n\rangle,
\label{eq:cohstate}
\eq
where $|n\rangle$ is an eigenstate of the plasmon number operator, viz. $a_{p}^{\dag}a_{p}|n\rangle=n_p|n\rangle$. The time dependent excitation probabilities $w_{\bf k_{1}}(t,t',t'')$ incorporate time-energy uncertainities arising from finite temporal integration boundaries on the RHS of (\ref{eq:w}).   Additional uncertainities may arise in (\ref{eq:Nplexpanded}) from integrations of pulse profiles over the final observation intervals $(t_0,t)$.

Expression (\ref{eq:cohstate})  bears analogy with the solutions for boson fields driven by external currents\cite{GlauberPR} or energetic particles\cite{Ritchie}, and the plasmonic polaron model\cite{FCaruso} in which the expression preceding each $\delta(\varepsilon-n\hbar\omega_{p})$ on the RHS of (\ref{eq:Nplexpanded}) gives the probability for $n$-plasmon excitation, in the present case caused by the electron component of laser pulse-induced polarization.\cite{satellites} 
Note also that in this nondispersive plasmon limit we readily obtain from (\ref{eq:N}), (\ref{eq:Ndirect}) and (\ref{eq:Nplexpanded}) the equivalence

\bq
{\cal N}^{pl}(N,t)\leftrightarrow \hbar\omega_{p}{\cal S}^{pl}(\varepsilon,t)
\label{eq:NS}
\eq
implementable through $\delta(N-n)\leftrightarrow\hbar\omega_{p}\delta(\varepsilon-n\hbar\omega_{p})$.

\subsection{Scattering boundary conditions limit of Eq. (\ref{eq:Nplexp})}
\label{sec:SBC}
%
\begin{figure}[tb] 
\rotatebox{0}{ \epsfxsize=8cm \epsffile{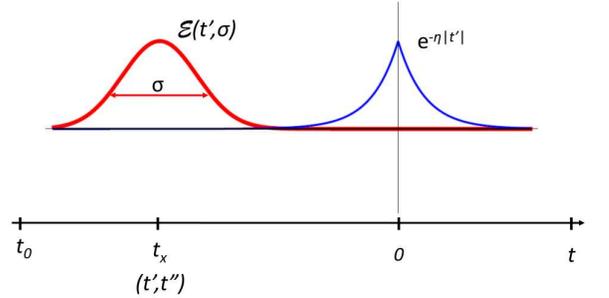}} 
\caption{Schematic illustration of the temporal behaviour of the external EM field amplitude ${\cal E}(\omega_x,t',\sigma_{x})$ (red curve) and $V^{e-pl}(t')=V^{e-pl}e^{-\eta|t'|}$ (blue curve, $\eta\rightarrow 0^{+}$) appropriate to the scattering boundary conditions for preparation of electrons in excited states $|{\bf k_1}\rangle$ above the Fermi level and their subsequent interaction with plasmons. The horizontal axis illustrates the hierarchy of times established in  expression (\ref{eq:hierarchy}).}
\label{SBC}
\end{figure}

 The scattering boundary conditions (SBC) limit of Eq. (\ref{eq:Nplexp}) is particularly illustrative as it allows to trace the energy conservation during the primary photoexcitation of electronic polarization and subsequent plasmon generation.   
To illustrate this we follow the standard collision theory approach\cite{GW} and explore the consequences of adiabatic switching on and off of the electron-plasmon interaction in Eq. (\ref{eq:allHint}) relative to the action of the pulse (\ref{eq:W}). Here we assume that the pulse centered at $t_x$ and with amplitude ${\cal E}(\omega_x,t,\sigma_{x})$ acts for sufficiently long preparation time $\sigma_{x}$ in the interval $t_0<t_x<-\eta^{-1}$ that is embedded in the much longer interval $(t_0\rightarrow -\infty,t\rightarrow\infty)$. This establishes the hierarchy among the various characteristic times\cite{Sakaue} appearing in expression (\ref{eq:Nplexp}), viz.
\bq
t_0\ll(t_x,t',t'')\ll-\eta^{-1} \ll 0\ll \eta^{-1}\ll t,  
\label{eq:hierarchy}
\eq
with $t_0\rightarrow -\infty$ and $t\rightarrow\infty$. This is sketched schematically in Fig. \ref{SBC}. Application of these SBC to definition (\ref{eq:DeltaText}) enables a standard substitution
\bq
 \Delta_{\bf \bar{k}_1,k_1}^{\bf q}(t,(t'+t'')/2)\left|_{\rm SBC}\right.\rightarrow 2\pi\delta(\epsilon_{\bf k_1}-\epsilon_{\bf \bar{k}_1}-\hbar\omega_{\bf q})
\label{eq:Deltaonshell}
\eq
into Eq. (\ref{eq:Nplexp}) which puts plasmon emission by primary photoexcited electron on the energy shell. This makes expressions in the first and second line on the RHS of (\ref{eq:Nplexp}) effectively factorizable in the time domain. Consequently, the integrals over $t'$ and $t''$ in the first line can be readily carried out to lead to an  expression describing the role of experimental apparatus in the energy conserving preparation of quasiparticles in the excited states $|{\bf k_{1}}\rangle$.

\subsubsection{Factorization of the "apparatus function"}
\label{sec:apparatus}

The compliance of expression (\ref{eq:Nplexp}) with conditions  (\ref{eq:hierarchy}) and $\omega_{x}^{-1}\ll \sigma_{x}<\eta^{-1}<t$  leads in (\ref{eq:Nplexpanded}) to the factorization of the "apparatus function" containing the effective fields 

\barr
{\cal A}_{\bf k_{1},k_{2}}(\omega_x,\sigma_{x})&=&|W_{\bf k_{1},k_{2}}|^{2}(1-n_{\bf k_{1}})n_{\bf k_{2}}\nonumber\\
&\times&
\left|\int_{-\infty}^{\infty}\frac{dt'}{\hbar}{\cal E}(\omega_x,t',\sigma_{x}) e^{-\frac{i}{\hbar}(\hbar\omega_{x} -\epsilon_{\bf k_{1}}+\epsilon_{\bf k_{2}})t'}\right|^{2},
\label{eq:apparatus}
\earr
which exhibits the property 
\bq
{\cal A}_{\bf k_{1},k_{2}}(\omega_x,\sigma_{x})\propto\sigma_{x}\delta(\hbar\omega_x-\epsilon_{\bf k_{1}}+\epsilon_{\bf k_{2}}).
\label{eq:calA}
\eq

Combining (\ref{eq:apparatus}) with  the second line in Eq. (\ref{eq:Nplexp}) the latter can be represented in the form of on-the-energy-shell spectrum ${\cal S}^{pl}(\varepsilon,t)$ with minimal time-energy uncertainities during both stages of primary photoexcitation of electronic polarization and sequential plasmon emission 

\begin{widetext}
\barr
\lim_{t\rightarrow\infty}{\cal S}^{pl}(\varepsilon,t)&=&\sum_{\bf k_{1},k_{2}} {\cal A}_{\bf k_{1},k_{2}}(\omega_x,\sigma_{x})\frac{1}{2\pi\hbar}\int_{-\infty}^{\infty}d\tau e^{\frac{i}{\hbar}\varepsilon\tau}\exp\left[-\sum_{\bf \bar{k}_1,q}w_{\bf \bar{k}_1,k_{1}}^{\bf q}(1-e^{-i\omega_{\bf q}\tau})\right]
\label{eq:NSBC}\\
&=&
\sum_{\bf k_{1},k_{2}} {\cal A}_{\bf k_{1},k_{2}}(\omega_x,\sigma_{x})e^{-w_{\bf k_{1}}}
\left[\delta(\varepsilon)+\sum_{\bf \bar{k}_1,q}w_{\bf \bar{k}_1,k_{1}}^{\bf q}\delta(\varepsilon-\omega_{\bf q})+\sum_{\bf \bar{k}_1,q;{\bf \bar{k}_{1}',q'}}w_{\bf \bar{k}_1,k_{1}}^{\bf q}w_{\bf \bar{k}_{1}',k_{1}}^{\bf q'}\delta(\varepsilon-\omega_{\bf q}-\omega_{\bf q'})+\cdots\right]\nonumber\\
\label{eq:NSBCexpanded}
\earr
\end{widetext}
Here
\bq
w_{\bf \bar{k}_1,k_{1}}^{\bf q}=\left|2\pi V_{\bf \bar{k}_1,k_{1}}^{\bf q}\delta(\epsilon_{\bf k_{1}}-\epsilon_{\bf \bar{k}_1}-\hbar\omega_{\bf q})\right|^{2}(1-n_{\bf \bar{k}_1}),
\label{eq:wscatt}
\eq
is a dimensionless transition probability in which the divergent square of the $\delta$-function can be regularized following the procedures appropriate to the treatments of explicitly time-dependent quantum transition probabilities.\cite{PhysRep,Messiah,FarisPRL} This is described in Sec.\,S1 of Ref. [\onlinecite{SM}]. 
As shown therein, from the on-the-energy-shell limit (\ref{eq:wscatt}) directly follows Fermi's golden rule expression for the differential transition rate describing electron transition  ${\bf k_1}\rightarrow {\bf \bar{k}_1}$ through the energy conserving emission of one plasmon of wavevector ${\bf q}$, viz. 
\bq
\Gamma_{\bf \bar{k}_1,k_1}=2\pi\sum_{\bf q}|V_{\bf \bar{k}_1,k_1}^{\bf q}|^2(1-n_{{\bf \bar{k}_1}})\delta(\epsilon_{\bf k_{1}}-\epsilon_{\bf \bar{k}_1}-\hbar\omega_{\bf q}).
\label{eq:Gammaplasdiff}
\eq
The saturation limits (\ref{eq:wscatt}) and (\ref{eq:Gammaplasdiff}) become effective already few femtoseconds after plasmon excitation.\cite{FarisPRL,LazicPRL}

The total transition rate per unit time, $1/\tau_{\bf k_1}=\Gamma_{\bf k_1}^{>}/\hbar$, for one plasmon excitation by an electron scattered from the primary photoexcited state $|{{\bf k}_{1}}\rangle$ {\it above} $E_F$ to all unoccupied final band states $|{\bf \bar{k}_1}\rangle$ satisfying energy conservation, is obtained from\cite{SS1975}
\bq
\Gamma_{\bf k_1}^{>}=\sum_{\bf \bar{k}_1}\Gamma_{\bf \bar{k}_1,k_1}=2\pi\sum_{\bf \bar{k}_1,q}|V_{\bf \bar{k}_1,k_1}^{\bf q}|^2(1-n_{{\bf \bar{k}_1}})\delta(\epsilon_{\bf k_{1}}-\epsilon_{\bf \bar{k}_1}-\hbar\omega_{\bf q}).
\label{eq:Gammaplas}
\eq
Here the summation over the final electron quantum numbers ${\bf \bar{k}_1}$  introduces the one-plasmon excitation threshold $\theta(\epsilon_{\bf k_{1}}-\hbar\omega_{\bf q}-E_F)$. The quantity ({\ref{eq:Gammaplas}) depends solely on the intrinsic properties of the electron-plasmon system.  Expressions (\ref{eq:Gammaplasdiff}) and (\ref{eq:Gammaplas}) will serve in Sec. \ref{sec:pumpingplasmon} for carrying out the calculations of primary hot electron decay caused by plasmon emision.  

It is notable that within the SBC the total energy conservation over the temporally nonoverlapping interactions $W(t')$ and $V^{e-pl}(t')$ in the interval $(t_0,t)$ is linked only via the primary excited quasiparticle energy $\epsilon_{\bf k_1}$. This is easily seen by combining energy conservations from the two nonoverlapping events of electron photoexcitation and plasmon emission. This yields for one-plasmon excitation the condition $\hbar\omega_x=\epsilon_{\bf \bar{k}_1}+\hbar\omega_{\bf q}-\epsilon_{\bf k_2}$ where $\epsilon_{\bf k_2}$ is the primary photoexcited hole energy. Furthermore, since in the preparation of momentum and energy resolved scattering experiments the primary quasiparticle momentum and energy are kept fixed, the "apparatus function" in (\ref{eq:NSBC}) obeying this condition will be effectively restricted to a single factorizable ${\bf k_1}$-component $\sum_{\bf k_{2}} {\cal A}_{\bf k_{1},k_{2}}(\omega_x,\sigma_{x})$. This leads to the formulation of the plasmon emission event in accord with the standard collision theory developed in Chs. 3-5 of Ref. [\onlinecite{GW}].

The coherent state related representation of (\ref{eq:NSBC}) is again most easily visualized in the simple nondispersive limit $\omega_{\bf q}=\omega_p$ employed in Sec. \ref{sec:nondispersive} which together with (\ref{eq:apparatus}) and (\ref{eq:w1}) yields a Poissonian distribution spectrum
\barr
\lim_{t\rightarrow\infty}{\cal S}^{pl}(\varepsilon,t)&=&\sum_{\bf k_{1},k_{2}} {\cal A}_{\bf k_{1},k_{2}}(\omega_x,\sigma_{x}) \nonumber\\
&\times&
\sum_{n=0}^{\infty}e^{-w_{\bf k_{1}}}\frac{\left(w_{\bf k_{1}}\right)^{n}}{n!}\delta(\varepsilon-n\hbar\omega_p),
\label{eq:NSBCcoh}
\earr
with
\bq
w_{\bf k_{1}}=\sum_{\bf \bar{k}_1,q}w_{\bf \bar{k}_1,k_1}^{\bf q}
\label{eq:w1}
\eq
In this limit expressions (\ref{eq:Psi}) and  (\ref{eq:cohstate}) tend to a superposition of electronic and quasistationary coherent plasmonic states that within the SBC acquire a compact form
\barr
|\Psi_{\rm SBC}\rangle&=& \sum_{{\bf k_{1},k_{2}},n=0}^{n=\infty} \sqrt{{\cal A}_{\bf k_{1},k_{2}}(\omega_x,\sigma_{x})}\nonumber\\
&\times& 
e^{-\frac{w_{\bf k_{1}}}{2}}\sqrt{\frac{\left(w_{\bf k_{1}}\right)^{n}}{n!}}|{\bf k_{1},k_{2}}\rangle|n\rangle.
\label{eq:cohscatt}
\earr
In the opposite case of temporally overlapping interactions $W(t')$ and $V^{e-pl}(t')$, the time dependences in the first and second line on the RHS of (\ref{eq:Nplexp}) do not factorize and simple closed form solutions are possible only for special pulse shapes (see Sec. \ref{sec:cohfinstate}).

\subsubsection{Truncated plasmonic coherent states}
\label{sec:truncated}

The plasmonic excitation spectrum given by the expression following the apparatus function on the RHS of (\ref{eq:NSBC}), (\ref{eq:NSBCexpanded}) and (\ref{eq:NSBCcoh}) is unitary by construction. However, as it was derived from second order cumulant expansion which approximates the higher order plasmon emission events by a succession of uncorrelated first order ones (cf. Appendix \ref{sec:ElectronPart}), the excitation threshold condition $\epsilon_{\bf k_1}-E_F\geq n\hbar\omega_p$ for $n$-plasmon emission ($n\geq 2$) does not explicitly appear in the third and higher order terms in the square brackets in (\ref{eq:NSBCexpanded}). To remedy this defficiency of second order cumulant approximation all the terms on the RHS of (\ref{eq:NSBCexpanded}) beyond the second one should be consecutively multiplied by the factors $\theta(\epsilon_{\bf k_{1}}-n\hbar\omega_{p}-E_{F})$ in order to remove excess weights that violate energy conservation. Since the thus modified spectrum violates unitarity it must be reunitarized with respect to the $n=0$ elastic line $e^{-w_{\bf k_{1}}}\delta(\varepsilon)$. One such procedure was outlined  in the last two paragraphs in Sec. III.C of Ref. [\onlinecite{chopoff}] and amounts to multiplying the  elastic line $\delta(\varepsilon)$ in expression (\ref{eq:NSBCexpanded}) by the sum of all removed ${\bf k_1}$-dependent weights 
\bq
\sum_{n=2}^{\infty}\sum_{\bf \bar{k}_1,q;...;\bar{k}_{1}',q'}w_{\bf \bar{k}_1,k_{1}}^{\bf q}\cdots w_{\bf \bar{k}_{1}',k_{1}}^{\bf q'}\theta(n\hbar\omega_p +E_F-\epsilon_{\bf k_1}).
\label{eq:chopoff} 
\eq
This procedure ensures that the sum of energies of excited plasmons can not exceed the absorbed photon energy and leaves the common Debye-Waller factor $e^{-w_{\bf k_{1}}}$ from (\ref{eq:NSBCexpanded}) unchanged. Besides, constraining $n$ by energy conservation transforms (\ref{eq:NSBCcoh}) into a sub-Poissonian distribution and the plasmonic wavefunction (\ref{eq:cohscatt}) into a truncated coherent state.

\subsection{Final plasmonic state in the limit of excitation by extremely short pulses}
\label{sec:cohfinstate}

\begin{figure}[tb] 
\rotatebox{0}{ \epsfxsize=9cm \epsffile{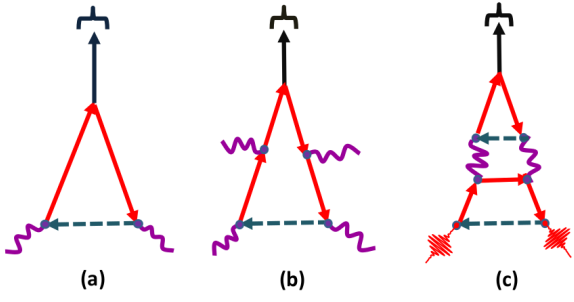}} 
\caption{Diagrammatic illustration of perturbative contributions to plasmon pumped electron yield following the procedures outlined in Refs. [\onlinecite{TimmBennemann}] and [\onlinecite{Caroli}]. Positive time is running upward. Sketch of Keldysh diagram for plasmonically induced electron yield in (a) quadratic response (analogous to 1PP yield), and in (b) quartic response (analogous to 2PP yield). Full red and dashed blue lines denote propagation of electrons and holes above and below the Fermi level, respectively. Purple wiggly lines indicate plasmons giving rise to electron excitation via electron-plasmon interaction matrix elements $V_{\bf \bar{k}_1,k_1}^{\bf q}$ denoted by blue dots. Vertical full black arrows symbolize time dependent electron yield captured by the electron counter. In the processes depicted in panels (a) and (b) the wiggly lines can be any real plasmon emanating from the primary photoinduced polarization vertices shown in Figs. \ref{Psi} and \ref{Psi2P} or their complex conjugates. 
(c) Illustration of the mechanism of plasmonic filter for electron yield obtained through temporal juxtaposition of the processes depicted in Fig. \ref{Psi} and the herein panel (a). This particular yield channel is open for $\hbar\omega_x>\hbar\omega_p>\phi$. }
\label{plasmonyield}
\end{figure}

In this example we consider the boundary conditions pertaining to extremely short driving field pulse duration in comparison with the switching on time of the electron-plasmon interaction, viz. the limit $1/\eta\gg\sigma_{x}$. This is the situation opposite to the one sketched in Fig. \ref{SBC}, so that now a narrow pulse envelope with $t_x$ close to $0$ is embedded within the adiabatic development of the electron-plasmon interaction governed by $e^{-\eta|t|}$. To explore the inferences of the very short pulse limit in a tractable calculation we follow Keldysh\cite{Keldysh2017} and assume a transient profile ${\cal E}(\omega_x,t,\sigma_{x})\rightarrow{\cal E}_0\sigma_{x}\delta(t-t_x)$. Using this Ansatz in the simple example (\ref{eq:Nplexpanded}) we find for $t>t_x$

\barr
{\cal S}^{pl}(\varepsilon,t)&=&
\left(\frac{{\cal E}_0\sigma_{x}}{\hbar}\right)^2
\sum_{\bf k_{1},k_{2}} |W_{\bf k_{1},k_{2}}|^{2}(1-n_{\bf k_{1}})n_{\bf k_{2}}\nonumber\\
&\times&
\sum_{n=0}^{\infty}e^{-w_{\bf k_{1}}(t)}\frac{\left[w_{\bf k_{1}}(t)\right]^{n}}{n!}\delta(\varepsilon-n\hbar\omega_{p}), 
\label{eq:Nplcoh}
\earr
with

\barr
w_{\bf k_{1}}(t)&=&
\sum_{\bf \bar{k}_1,q}(1-n_{\bf \bar{k}_1})\left|V_{\bf \bar{k}_1,k_{1}}^{\bf q}\right|^{2}\nonumber\\
&\times&
2\left(\frac{1-\cos[(\epsilon_{\bf k_{1}}-\epsilon_{\bf \bar{k}_1}-\hbar\omega_{p})(t-t_x)]}{(\epsilon_{\bf k_{1}}-\epsilon_{\bf \bar{k}_1}-\hbar\omega_{p})^{2}}\right).
\label{eq:v}
\earr
Observe in (\ref{eq:Nplcoh}) the modified "apparatus function" $({\cal E}_0\sigma_{x}/\hbar)^2 |W_{\bf k_{1},k_{2}}|^{2}(1-n_{\bf k_{1}})n_{\bf k_{2}}$ relative to expression (\ref{eq:apparatus}). Also notable is the absence of a monochromatic energy conservation in the primary excitation process [otherwise arising from the upper line in (\ref{eq:Nplexp})] which is here triggered by the white spectrum of $\delta$-function pulse. 

In the case of subcycle pump pulses the physically relevant limit of (\ref{eq:Nplcoh}) is for $t\rightarrow\infty$.\cite{Keldysh2017} Employing this limit expression (\ref{eq:v}) tends to (\ref{eq:wscatt}) in which the $\delta$-functions should be converted to nondiverging expressions via the appropriate regularization procedures described in Sec.\,S1 of Ref. [\onlinecite{SM}]. Hence, both extreme regimes of optical excitation considered in the previous and present subsections support the formation of coherent plasmonic states (\ref{eq:cohstate}) with a common distributional probability (\ref{eq:wscatt}). 
However, the distinct temporal boundary conditions imposed on the optical pumping processes give rise to different "apparatus functions" in (\ref{eq:NSBCcoh}) and (\ref{eq:Nplcoh}), and thereby to the pulse-specific weights of the respective plasmonic spectra $S^{pl}(\varepsilon,t)$.

\section{Secondary hot plasmon-driven pumping of nonlinear electron yield}
\label{sec:plasmonpump}

The presence of a distribution of real plasmons in the system, either in the form of "coherent states", (\ref{eq:cohstate}) and (\ref{eq:cohscatt}), or a more general one, may have interesting implications on the dynamics of electronic excitations in the valence and conduction bands. Like the photons of applied EM field, real plasmons propagating in the system may also act, via the interaction $V^{e-pl}$ in (\ref{eq:Hsyst}), as a secondary pump field for excitation of electrons to states above the vacuum level. These electrons would emanate from the system with kinetic energies solely determined by the multiples of plasmon energy.  This is illustrated by the diagrams (a) and (b) in Fig. \ref{plasmonyield}.  The essential  difference between the plasmon and photon drivings appears first in the meaning of wiggly lines denoting the exciting boson fields (plasmon vs. photon), and second, in the interaction vertices describing their coupling to electrons ($V_{\bf \bar{k}_1,k_1}^{\bf q}$ vs. $W_{\bf k_1,k_2}$, respectively). Another difference arises in temporal constraints on the two boson fields. In the case of photons they are imposed by the modulation ${\cal E}(\omega_x,t,\sigma_{x})$ of ultrashort pulses. By contrast, once the "hot" plasmons have been excited by the primary photoinduced electronic polarization they can be regarded as a steady continuous wave (cw) field whose action on electrons is limited only by the plasmon lifetime.\cite{Marini2002} 

The transient secondary plasmon-induced electron population  $\langle N_{\bf \bar{k}}(t)\rangle$ in an outgoing state $|{\bf \bar{k}}\rangle$ can be most directly demonstrated by taking the mean value of the ${\bf \bar{k}}$-resolved electron occupation operator $\hat{N}_{\bf \bar{k}}=c_{\bf \bar{k}}^{\dag}c_{\bf \bar{k}}$ over the quasistationary multiplasmonic state $|\Psi(0)\rangle$ reached at $t=0$ (which requires $t_x+\sigma <0$) and subsequently developed by the long time limit $t\gg 0$ of the evolution operator  (\ref{eq:Usyst}), viz. 
\bq
\langle N_{\bf \bar{k}}(t)\rangle=\langle\Psi(0)|U^{syst}(t,0)^{\dag}\hat{N}_{\bf \bar{k}}U^{syst}(t,0)|\Psi(0)\rangle
\label{eq:J_k}
\eq
 Now, in analogy with the $m$PP formalism (cf. Eqs. (4) and (10) in Ref. [\onlinecite{encyclopedia}]) expanding $U^{syst}(t,0)$ in powers of $V^{e-pl}$ gives the quadratic, quartic, etc. response expressions for plasmon-driven electron yields described next.

\subsubsection{One plasmon-assisted electron yield}
\label{sec:1yield}

We shall formulate the rate of creation of hot electron population in bulk bands through one plasmon-driven processes within the quadratic response approach adopted from the theory of cw-driven 1PP.\cite{Ashcroft,Mahan1PP,Caroli} This is illustrated diagrammatically in Fig. \ref{plasmonyield}(a) and can be studied in two complementary usages of state-to-state resolved electron excitation rates

\bq
\gamma_{\bf k',k}^{(\hbar\omega_p)}=2\pi\sum_{\bf q}|V_{\bf k',k}^{\bf q}|^{2}n_{\bf k}(1-n_{\bf k'})\delta(\epsilon_{\bf k'}-\epsilon_{\bf k}-\hbar\omega_{\bf q}).
\label{eq:gamma1resolved}
\eq
First, one may be interested as how a particular initial electron state $|{\bf k}\rangle$ below $E_F$ contributes to the total hot electron production in the states $|{\bf k'}\rangle$. This would correspond to the CIS mode and the pertinent electron excitation rate is given by $\gamma_{\bf k}^{{\rm CIS}(\hbar\omega_p)}/\hbar$ where 

\bq
\gamma_{\bf k}^{{\rm CIS}(\hbar\omega_p)}=\sum_{\bf k'}\gamma_{\bf k',k}^{(\hbar\omega_p)}.
\label{eq:GammaCIS}
\eq
Alternatively, one may be interested in how much the manifold of initial states $\{|\bf k\rangle\}$ gives rise to plasmon-induced excitation of electrons into a particular state $|{\bf k'}\rangle$ above $E_F$. The corresponding constant final state (CFS) rate 
is given by $\gamma_{\bf k'}^{{\rm CFS}(\hbar\omega_p)}/\hbar$ where 
\bq
\gamma_{\bf k'}^{{\rm CFS}(\hbar\omega_p)}=\sum_{\bf k}\gamma_{\bf k',k}^{(\hbar\omega_p)}.
\label{eq:GammaCFS}
\eq
Then, the expression
\bq
\gamma^{(\hbar\omega_p)}=\sum_{\bf k}\gamma_{\bf k}^{{\rm CIS}(\hbar\omega_p)}=\sum_{\bf k'}\gamma_{\bf k'}^{{\rm CFS}(\hbar\omega_p)},
\label{eq:gamma1tot}
\eq
gives the total electron excitation rate induced by one plasmon absorption.

The transitions rates $\Gamma_{\bf k_1}^{>}$, $\gamma_{\bf k}^{{\rm CIS}(\hbar\omega_p)}$ and $\gamma_{\bf k'}^{{\rm CFS}(\hbar\omega_p)}$ are appropriate for description of electron-plasmon scattering dynamics subject to SBC defined in (\ref{eq:hierarchy}).  
Note also that the excitation rates (\ref{eq:GammaCIS}) or (\ref{eq:GammaCFS}), on the one hand, and (\ref{eq:Gammaplas}) on the other hand, are not simple time reversals of each other because in the discussed experiment they involve different subspaces of initial and final electron states.

\subsubsection{Two plasmon-assisted electron yield}
\label{sec:2yield}

Multiplasmon-assisted electron yields require the existence of more than one real plasmon in the system. Once the real plasmons have been excited their annihilation  can give rise to $2\hbar\omega_p$-features presented in Fig. \ref{AllAg}. This process can be visualized as a quartic response of the electron system to the plasmon field and is illustrated diagrammatically in Fig. \ref{plasmonyield}(b).  Its description can be brought into a full analogy with the earlier formulated theory of 2PP processes presented in Sec. 2.1 of Ref. [\onlinecite{UebaGumhalter}], but with real plasmons instead of photons acting as the pump and probe fields. Hence, in the  long time limit expression (\ref{eq:J_k}) produces the excitation rate describing the total energy-resolved two plasmon-induced electron yield
\begin{widetext}
\bq
P(\epsilon)=\sum_{\bf k'',k}\left|2\pi\sum_{\bf k',q',q}\frac{V_{\bf k'',k'}^{\bf q'}V_{\bf k',k}^{\bf q}}{\epsilon_{\bf k'}-\epsilon_{\bf k}-\hbar\omega_{\bf q}+i\delta}\delta(\epsilon_{\bf k''}-\epsilon_{\bf k}-\hbar\omega_{\bf q}-\hbar\omega_{\bf q'})                 \right|^{2}\delta(\epsilon-\epsilon_{\bf k''}),
\label{eq:2plasPE}
\eq
\end{widetext}
where ${\bf k}$, ${\bf k'}$ and ${\bf k''}$ denote the quantum numbers of initial occupied, and intermediate and final unoccupied electron states, respectively. Expression on the RHS of (\ref{eq:2plasPE}) considerably simplifies in the case of weakly dispersive plasmons for which $\omega_{\bf q}\simeq\omega_p$.   

 The most notable feature of the above described plasmon pumping mechanism is its monochromaticity fixed at the multiples of plasmon frequency  $n\omega_p$, irrespective of the value of primary photon pump frequency $\omega_x$. Thus, as long as $\omega_x\geq \omega_p$ in 2PP the secondary plasmon-driven pumping of electronic excitations may become effective for the times $t>t_x+\sigma_{x}$. This enables the non-Einsteinian plasmonic CIS mode of photoelectron yield from the Fermi level at kinetic energies $2\hbar\omega_p-\phi>0$.  

\subsubsection{Plasmonic filter of electron yield}
\label{sec:filter}

The locking of electron emission energy at the plasmon energy can take place also in a temporal succession of the primary photon-induced plasmon generation shown schematically in Fig. \ref{Psi}, and the secondary plasmon-driven pumping of electronic polarization shown in Fig. \ref{plasmonyield}(a). One of the lowest order contributions (i.e. quadratic in photon and plasmon fields) to this combined process is sketched in Fig. \ref{plasmonyield}(c). This mechanism becomes effective provided the condition  $\hbar\omega_x>\hbar\omega_p>\phi$ holds and the limit of electron detection times exceeding the characteristic times of the system is reached. The latter is required  because here plasmons appear in the intermediate states past the primary optically induced  electronic transitions and not as a "preexistent" distribution of real excitations already available in (\ref{eq:cohscatt}) for generation of electron yield via the processes  sketched in panels (a) and (b) of Fig. \ref{plasmonyield}. Hence, for the plasmonic energy filter to be effective the interval  between the plasmon creation times [the times of lower plasmon vertices in Fig. \ref{plasmonyield}(c)] and the electron detection time (the time of topmost vertex in the same plot) must be sufficiently long to allow for establishment of energy conservation between the successive processes. Most importantly, this does not imply the resonance condition $\hbar\omega_{x}\simeq \hbar\omega_{p}$, as dominantly discussed in plasmonics,\cite{Linic,Boriskina,Narang2016,Nordlander15,deAbajo16} but only the requirement $\hbar\omega_{x}\geq \hbar\omega_{p}$.\cite{RefCaroli} 
 The same applies also to higher order processes [e.g. involving two-photon pumping succeeded by the subprocess illustrated in Fig. \ref{plasmonyield}(b)] albeit in a more complicated fashion. Thus, for CIS mPP from the states near the Fermi surface the electron yield may exhibit secondary peak structures located at multiples of the plasmon energy $n\hbar\omega_p$ above $E_F$ and their interpretation is possible only beyond the standard Einstein picture of photoeffect quantified through Eq. (\ref{eq:Einstein}). However, the coherence of pumping processes producing the electron yield can be quite different for photon and plasmon driven emissions.

\section{Results and discussion}
\label{sec:results}

In Refs.\,[\onlinecite{MarcelPRL,AndiNJP,ACSPhotonics}] we have presented exhaustive descriptions of the Einsteinian one-electron aspects of 2PP spectra from three low index surfaces of Ag that is consistent with the common understanding of such processes. In this section we focus on applications of concepts and results obtained in Secs.\,\ref{sec:System}-\ref{sec:plasmonpump} to study the concomitantly discovered non-Einsteinian plasmon-induced electron yield from Ag surfaces reported and heuristically interpreted in Ref.\,[\onlinecite{ACSPhotonics}]. Accordingly, we shall consider primary electronic excitations induced by photon energies in the range shown on the horizontal axes in Fig.\,\ref{AllAg}. Such low photon energies limit the truncated coherent plasmonic states generated in quadratic response by one photon-induced electronic polarization to contain maximum one real plasmon [cf. discussion preceding Eq. (\ref{eq:chopoff})]. Multiplasmon states are then generated in higher order response, i.e. quartic, etc (cf. Fig. \ref{Psi2P}). Also, as the on-the-energy-shell limits of transition rates (\ref{eq:Gammaplasdiff}) and (\ref{eq:Gammaplas}) are established in few femtoseconds past the instant of plasmon creation\cite{FarisPRL,LazicPRL,PSS2012} we shall henceforth consider only the results from this limit.

\subsection{Band structure and dielectric properties of Ag}
\label{sec:bandstructure} 

Calculations of the various quasiparticle excitation rates (\ref{eq:Gammaplas}), (\ref{eq:GammaCIS}), (\ref{eq:GammaCFS}) and (\ref{eq:2plasPE}) require realistic input quantities for the underlying minimal plasmon model. This specifically concerns the Ag electronic band structure and the ensuing dielectric properties.  
Figure \ref{preliminaries} illustrates the computed  electronic band structure over the first Brillouin zone (BZ) of bulk Ag and the thereof derived dielectric function (for details of ensuing calculation procedures see Ref. [\onlinecite{ACSPhotonics}] and Sec. S2 of Supplementary Material\cite{SM} building on Refs. [\onlinecite{bib:qe1,bib:pseudo,bib:pbe,bib:novko1,bib:novko2,bib:caruso1,bib:epw,bib:caruso2,bib:wan90,bib:autowan90,bib:gw1,bib:gw2}]). The extrema and kinks in the band dispersion curves around the endpoints of arrows denoting interband transitions (ib) in Fig. \ref{preliminaries}(a) give rise to maxima in the electronic density of states (DOS). Such maxima can strongly affect optical and plasmonic transition matrix elements in specific energy regions (see Sec. S2.C in Ref. [\onlinecite{SM}]) and thereby also the excitation rates (\ref{eq:Gammaplas}), (\ref{eq:GammaCIS}) and (\ref{eq:GammaCFS}).  

\begin{figure*}[tb] 
\rotatebox{0}{ \epsfxsize=14cm \epsffile{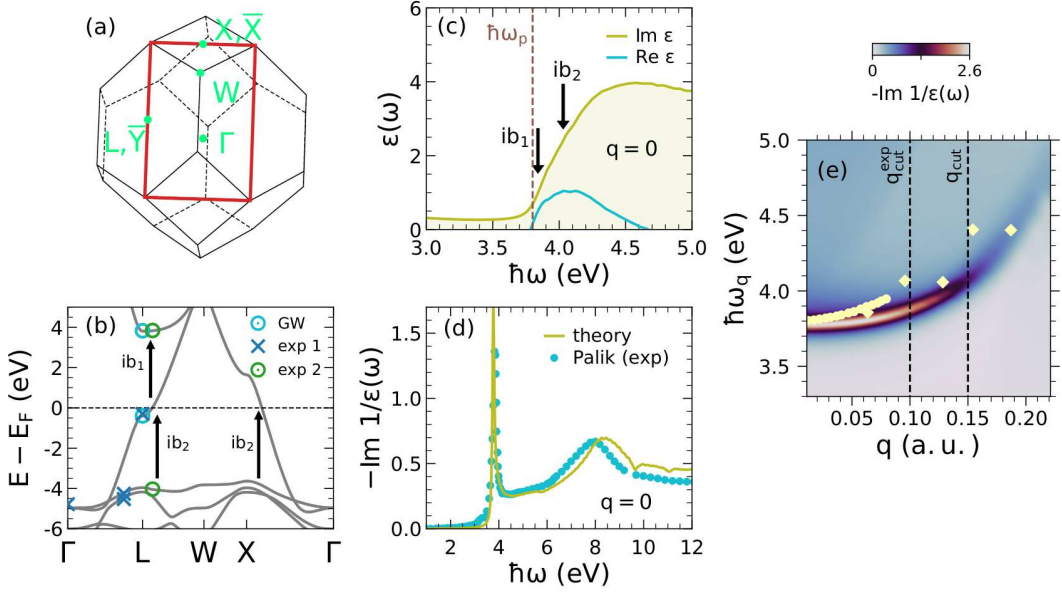}} 
\caption{(a): First Brillouin zone of \emph{fcc} Ag lattice. Green dots denote the high symmetry points of the zone and the red rectangle its section in the (110) plane. (b): Band structure of Ag along the high symmetry points in the first Brillouin zone computed within the GW-framework. The most prominent interband transitions (ib) in the energy range $\sim$ 4 eV are denoted by vertical arrows.  Earlier GW results (circles) are taken from Ref. [\onlinecite{Louie15}]. Experimental points are from Refs. [\onlinecite{exp1}] and [\onlinecite{exp2}]. (c): Real and imaginary parts of the RPA dielectric function for ${\bf q=0}$ calculated for the band structure displayed in (a).  (d): Comparison of the thereof calculated loss function with the experimental data from Ref. [\onlinecite{Palik}]. (e): Bulk plasmon dispersion in Ag calculated in RPA for the band structure shown in (b). Circles and diamonds denote experimental points of plasmon dispersion.[\onlinecite{lindau,zacharias}]}
\label{preliminaries}
\end{figure*}

\subsection{Hot plasmon generation}
\label{sec:pumpingplasmon}

Figure \ref{wk1}(a) shows the locations of ${\bf k}$-states along the high symmetry directions in the first BZ of Ag from which the decay of hot electrons gives rise to significant plasmon generation rates $\Gamma_{\bf k}^{>}$ defined in Eq. (\ref{eq:Gammaplas}). Depending on the initial electron ${\bf k}$-state these values can be as large as $\sim 120$ meV.  This implies that plasmons can be excited already within $5-6$ femtoseconds past the primary optical polarization excitation by the pump pulse, i.e. well within the duration of laser pulses used in experiments.\cite{MarcelPRL,ACSPhotonics} Analogous calculations of  $\Gamma_{\bf k}^{<}$ corresponding to plasmon generation by holes scattered from the initial $s,p$-band states $|{\bf k}\rangle$ extending to more than 4 eV below $E_F$ give negligible values. This finding provides {\it a fortiori} support to the neglect of plasmon emission by holes photoinduced in $s,p$-bands that was assumed in the derivation of expression (\ref{eq:Ndiag}). Good agreement between the angular integrated $\Gamma_{\bf k}^{>}$ shown in Fig. \ref{wk1}(b) and the analogous GW-derived quantity\,\cite{SM} justifies the use of minimal electron-plasmon model in the present calculations.

\subsection{$\hbar\omega_p$-yield}
\label{sec:plasmon1}
%

\begin{figure}[tb] 
\rotatebox{0}{ \epsfxsize=8cm \epsffile{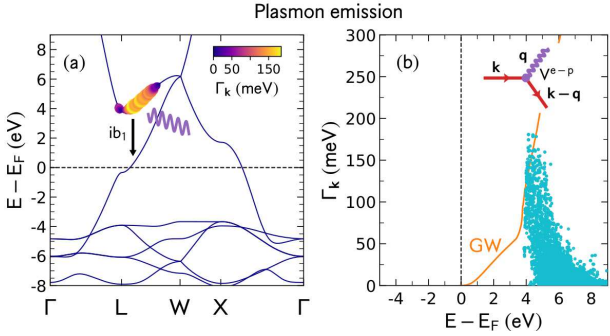}} 
\caption{ (a): Electronic band structure along the high symmetry points of the first  Brillouin zone in Ag. The magnitude of the band- and momentum-resolved photoexcited quasiparticle decay rate due to one plasmon emission $\Gamma_{\bf k}$, Eq. (\ref{eq:Gammaplas}), is quantified by color bar in the inset. 
 (b): Angle-integrated $\Gamma_{\bf k}$ as a function of energy. Note the threshold at $\sim E_F+\hbar\omega_p$ and very low values below $E_F$ and the $d$-band edge. Yellow line denotes the result of analogous GW-based calculation which  incorporates the full linear electronic density response of the system.}
\label{wk1}
\end{figure}

(i) {\it Bulk transitions.} Figure \ref{hotelectron} illustrates the rates $\gamma_{\bf k}^{{\rm CIS}(\hbar\omega_p)}$ and $\gamma_{\bf k'}^{{\rm CFS}(\hbar\omega_p)}$ of creation of hot electron population in bulk Ag through the plasmon driven processes sketched in Fig. \ref{plasmonyield}(a) (for methods of calculation see Sec.\,S2 in Ref.\,[\onlinecite{SM}]).   Panel (a) shows the locations of  ${\bf k}$-states in the BZ for which $\gamma_{\bf k}^{{\rm CIS}(\hbar\omega_p)}$ describing electron excitation from a fixed occupied state $|{\bf k}\rangle$ below $E_F$ to the states $|{\bf k'}\rangle$ above $E_F$ is large. Panel (b) shows equivalent information on the energy scale of  the occupied bands. Here two domains with maximum electron excitation rates are clearly discernible. They correlate with the enhanced DOS at the upper edges of occupied $d$-bands and the very narrow energy window just below $E_F$.\cite{Phillips}  The latter case of plasmon-induced electron excitation from $E_F$ which promotes  electrons to the states of energy $\sim E_F+\hbar\omega_p$ is overwhelmingly dominant and offers explanation of the $\hbar\omega_p$ peak in the CIS 1PP spectra  of Ref. [\onlinecite{Horn}]. This also supports the conjectures of dominant plasmon-induced electron excitation from $E_F$ made in Ref. [\onlinecite{Hopfield1965}].
  Panel (c) of Fig. \ref{hotelectron} shows a complementary picture in which $\gamma_{\bf k'}^{{\rm CFS}(\hbar\omega_p)}$ describing electron excitation from all occupied states $|{\bf k}\rangle$ below $E_F$ to a fixed final state $|{\bf k'}\rangle$ above $E_F$ is maximum. Again, two distinct types of plasmon-induced interband transitions are dominant. The first are from the upper edge of $d$-band to $E_F$, and thus unobservable in photoemission. The second are from $E_F$ to the states with energy $\sim E_F+\hbar\omega_p$, and thus observable in the experiment described in Ref. [\onlinecite{Horn}].

The results presented in Figs. \ref{preliminaries}-\ref{hotelectron} above quantify  the sequence (\ref{eq:sequence}) leading to the energy-locked excitation of electrons to energy $\hbar\omega_p$ above the Fermi level of the irradiated system. 
Essentially, this sequence appears on the LHS, and its conjugate on the RHS, of the diagram shown in Fig. \ref{plasmonyield}(c). Here the vertex encompassing one incoming and one outgoing red electron line and one wavy plasmon line can be identified with the plasmaron vertex.\cite{Chis}  Note that the two-hole state $|\rm{polarization}''\rangle$ in (\ref{eq:sequence}) is mapped in the appearance of two hole dashed lines in Fig. \ref{plasmonyield}(c). 
The measures of durations of pumping of the involved excited states are $\sigma_{x}$, $\hbar/\Gamma_{\bf k}^{>}$ and $\hbar/\gamma_{\bf k'}^{CFS}$, respectively. The occurence of $\hbar\omega_p$ component of 1PP spectrum from Ref. [\onlinecite{Horn}], that can not be explained by relations (\ref{eq:Einstein}) or (\ref{eq:satellite}), is now interpretable by the scenario (\ref{eq:sequence}) and supported by the results of calculations presented in Figs. \ref{preliminaries}-\ref{hotelectron}.

\begin{figure}[tb] 
\rotatebox{0}{ \epsfxsize=8cm \epsffile{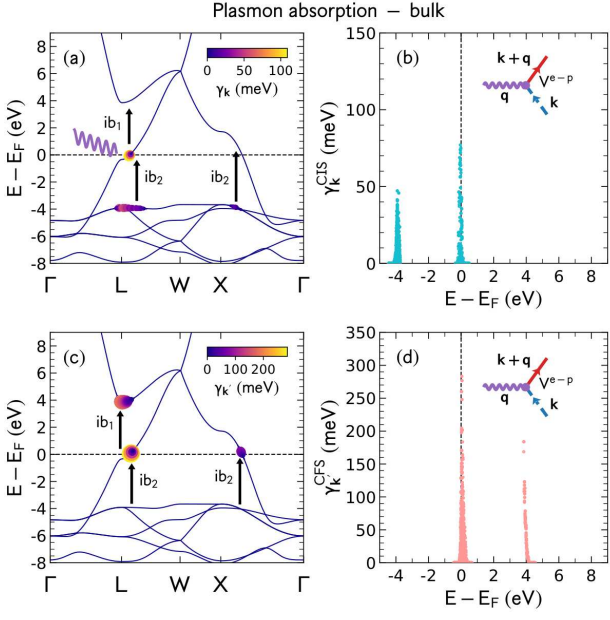}} 
\caption{(a): Band structure of Ag with electronic transitions ib2 and ib1 contributing to CIS resolved rates $\gamma_{\bf k}^{{\rm CIS}(\hbar\omega_p)}$ [Eq. (\ref{eq:GammaCIS})] for plasmon-induced hot electron generation processes sketched in Fig. \ref{plasmonyield}(a). The magnitudes of CIS and CFS transition rates are quantified by  color bars in the insets. (b):  Angle-integrated $\gamma_{\bf k}^{{\rm CIS}(\hbar\omega_p)}$ shown on the energy scale. The CIS amplitudes for transitions from the occupied states closely below $E_F$ to unoccupied bands above $E_F$ are dominant. (c): Electronic transitions ib2 and ib1 contributing to CFS resolved plasmon-induced hot electron generation rates $\gamma_{\bf k'}^{{\rm CFS}(\hbar\omega_p)}$ [Eq. (\ref{eq:GammaCFS})] from all initial to a fixed final ${\bf k}$-state at $\sim E_F$ and final ${\bf k'}$-state at $\sim E_F+\hbar\omega_p$, respectively. (d):  Angle integrated $\gamma_{\bf k'}^{{\rm CFS}(\hbar\omega_p)}$ shown on the energy scale. Here the dominant amplitudes are for transitions to $E_F$ and $\sim E_F+\hbar\omega_p$. }
\label{hotelectron}
\end{figure}

(ii) {\it Bulk and surface transitions.} To investigate and illustrate the effects of the surface on plasmon-assisted electron yield we carry out calculations analogous to those in (i) above by projecting the bulk band states onto the Ag(110) crystal plane (for calculational details see Sec.\,S2 E in Ref.\,[\onlinecite{SM}]).    
Figure \ref{hotelectron(110)}(a) illustrates as how due to the breakdown of translational symmetry perpendicular to the surface new states appear in the gaps of the bulk band structure. The concomitant relaxation of momentum conservation perpendicular to the surface opens new channels of hot electron excitation by plasmon absorption. This gives rise to the enhancements of hot electron excitation rates $\gamma_{\bf k}^{CIS(\hbar\omega_p)}$ and $\gamma_{\bf k'}^{CFS(\hbar\omega_p)}$, as can be deduced from the intensities of $\gamma_{\bf k'}^{CFS(\hbar\omega_p)}$ shown in Fig. \ref{hotelectron(110)}(a). The contributions of surface-induced channels to plasmon-assisted hot electron generation can be best visualized from the comparison of angular integrated surface enhanced $\gamma_{\bf k'}^{CFS(\hbar\omega_p)}$ shown in Fig. \ref{hotelectron(110)}(b) with its bulk counterpart shown in Fig. \ref{hotelectron}(d). The different vertical scales in the two plots reveal different magnitudes of the respective quantities. Furthermore, the surface-induced channels give rise to the nonvanishing intensity of $\gamma_{\bf k'}^{CFS(\hbar\omega_p)}$ also in the intermediate interval $(E_F<\epsilon<E_F+\hbar\omega_p)$ but the largest contributions at $E_F$ and $E_F+\hbar\omega_p$ persist, in accord with experimental trends observed for the other low index Ag surfaces.\cite{Horn}
The comparison between bulk and surface results nicely illustrates how surface effects can be of paramount importance in descriptions of hot carrier generation in plasmonic materials.

\begin{figure}[tb] 
\rotatebox{0}{ \epsfxsize=7.5cm \epsffile{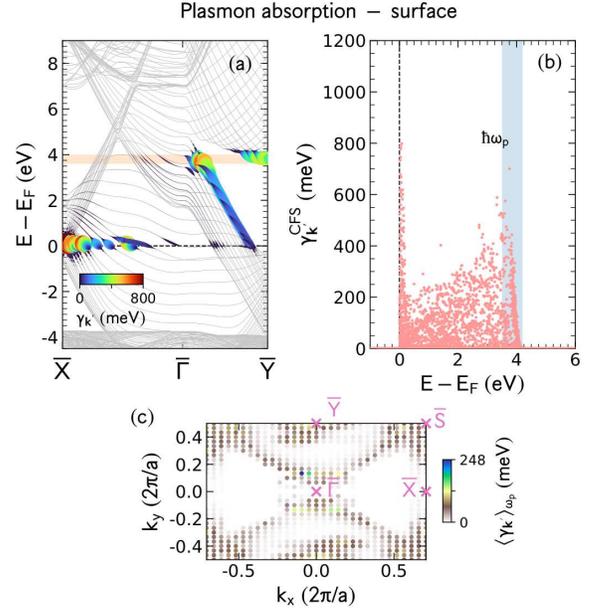}} 
\caption{(a): Band structure of Ag across the (110)-surface projected bulk Brillouin zone showing the intensities of surface enhanced $\gamma_{\bf k'}^{{\rm CFS}(\hbar\omega_p)}$ quantified by the color bar.
(b): Angle integrated surface enhanced $\gamma_{\bf k'}^{{\rm CFS}(\hbar\omega_p)}$ from (a) shown on the energy scale. Shaded area spans the width of $E_F\rightarrow E_F+\hbar\omega_p$ plasmon-induced electronic transitions. Notable  is the surface induced enhancement of all displayed quantities relative to the purely bulk induced values shown in Fig. \ref{hotelectron}. (c): Isoenergetic distribution of surface enhanced $\gamma_{\bf k'}^{{\rm CFS}(\hbar\omega_p)}$ averaged around $\epsilon_{\bf k'}\simeq E_F+\hbar\omega_p$[\onlinecite{averaged}] and shown across the projected BZ ($a=4.06\,{\rm \AA}$). The highest intensities are around $\bar{\Gamma}$, $\bar{\rm Y}$ and $\bar{S}$ points.  }
\label{hotelectron(110)}
\end{figure}

\subsection{$2\hbar\omega_p$-yield}
\label{sec:plasmon2}

The 2PP features shown in Fig. \ref{AllAg}, recorded with pulse frequencies $\omega_x$ obeying $\omega_p\leq\omega_x\leq 2\omega_p$, can not be visualized within a unique excitation sequence of the type (\ref{eq:sequence}) because  the excited states giving rise to the $2\hbar\omega_p$-yield may be reached via several different excitation paths involving two plasmons. This is symbolically represented by Fig. \ref{Psi2P} after exclusion of the plasmon lines emanating from the holes.  
However, concepts analogous to those employed in the previous subsection to interpret the $\hbar\omega_p$-yield  may be extended to interpret the $2\hbar\omega_p$ features as well. The point of departure is the polarization-induced formation of plasmonic coherent states governed by $\Gamma_{\bf k_1}^{>}$ illustrated in Fig. \ref{wk1}. The thus generated real plasmons may give rise to primary and secondary plasmon pumping over virtual and real intermediate states, i.e. nonresonant and resonant intermediate excitation paths. 
%

\begin{figure}[tb] 
\rotatebox{0}{\epsfxsize=8cm \epsffile{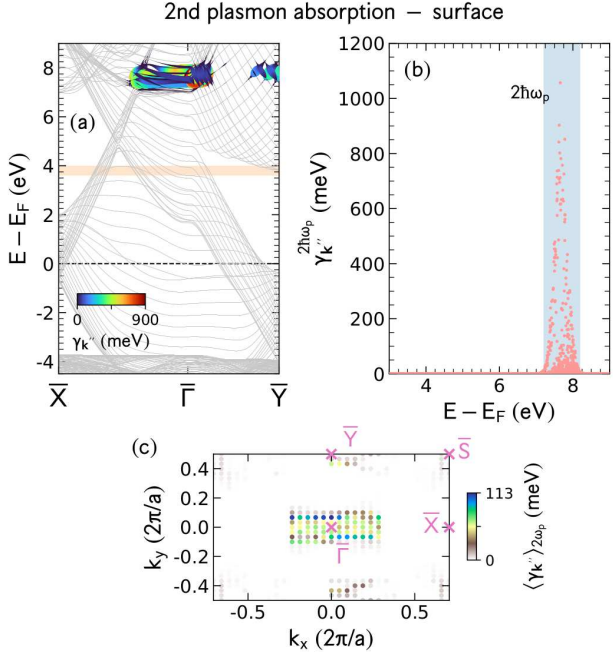}} 
\caption{(a):  Band structure of Ag projected onto (110) surface showing the points with maximal intensities of $\gamma_{\bf k''}^{CFS(2\hbar\omega_p)}$ (see color bar for computed values). The second plasmon-induced electronic transitions start from resonant intermediate states lying in the orange shaded energy window at $\sim E_F+\hbar\omega_p$ that have been occupied in one-plasmon absorption processes illustrated in panels (c) and (d) of Fig. \ref{hotelectron(110)}.
(b): Angle integrated $\gamma_{\bf k''}^{CFS(2\hbar\omega_p)}$ from (a) shown on the energy scale. Shaded area straddles the width of plasmonically-induced electronic transitions to $\sim E_F+2\hbar\omega_p$.  (c): Isoenergetic distribution of $\gamma_{\bf k''}^{{\rm CFS}(2\hbar\omega_p)}$ averaged around $\epsilon_{\bf k''}\simeq E_F+2\hbar\omega_p$[\onlinecite{averaged}] and shown across the projected BZ ($a=4.06\,{\rm \AA}$). The accumulation of highest intensities is in the vicinity of $\bar{\Gamma}$ and $\bar{\rm Y}$ points. }
\label{hotelectron2(110)}
\end{figure}

In the calculations of the one plasmon-driven electronic excitations reported in subsection \ref{sec:plasmon1} above we have identified a rich phase space of resonant states available for the one-plasmon pumping stage [cf. Fig. \ref{hotelectron}(c)]. 
These states also support resonant intermediate channels in the two-plasmon driven pumping of electrons from $E_F$. Since dominant contributions to dissipative processes   come from on-the-energy-shell transitions we shall henceforth consider only the manifold of sequential electron excitation paths via the resonant intermediate states.\cite{KeldyshIonization} Therefore, the second plasmonically induced electronic excitation from an intermediate resonant state $|{\bf k'}\rangle$ to a final state $|{\bf k''}\rangle$ is characterized by the state-to-state excitation rate 
\bq
\gamma_{\bf k'',k'}^{(2\hbar\omega_p)}=2\pi\sum_{\bf q'}|V_{\bf k'',k'}^{\bf q'}|^{2}n_{\bf k'}(1-n_{\bf k''})\delta(\epsilon_{\bf k''}-\epsilon_{\bf k'}-\hbar\omega_{\bf q'}).
\label{eq:gamma2resolved}
\eq
Here the energies $\epsilon_{\bf k'}\simeq E_F+\hbar\omega_p$ of the intermediate resonant states $|{\bf k'}\rangle$ occupied after the first plasmon absorption lie  within the orange horizontal stripe shown in Fig. \ref{hotelectron2(110)}(a).
Expression (\ref{eq:gamma2resolved}) is an analog of the one-plasmon induced transition rate $\gamma_{\bf k',k}^{(\hbar\omega_p)}$ defined in (\ref{eq:gamma1resolved}). For the second plasmon absorption processes the CFS rate reads
\bq
\gamma_{\bf k''}^{{\rm CFS}(2\hbar\omega_p)}=\sum_{\bf k'}\gamma_{\bf k'',k'}^{(2\hbar\omega_p)}.
\label{eq:gamma2}
\eq
This description is akin to treating the second step of 2PP proposed in Ref. [\onlinecite{Freericks}], but with hot plasmons instead of photons playing the role of probe field in the middle vertices in Fig. \ref{plasmonyield}(b). 

Panel (a) in Fig. \ref{hotelectron2(110)} shows the values of excitation rates $\gamma_{\bf k''}^{{\rm CFS}(2\hbar\omega_p)}$ (\ref{eq:gamma2}) for the second plasmon-induced electronic transitions in the projection of bulk BZ of Ag onto the (110) crystal surface. Here all transitions commence in the ${\bf k'}$-states within the orange stripe at the energy $\epsilon_{\bf k'}\simeq E_F+\hbar\omega_p$ occupied in one-plasmon absorption processes described by $\gamma_{\bf k'}^{{\rm CFS}(\hbar\omega_p)}$, and terminate in the ${\bf k''}$-states with $\epsilon_{\bf k''}\simeq E_F+2\hbar\omega_p$. Notable are the magnitudes of $\gamma_{\bf k''}^{{\rm CFS}(2\hbar\omega_p)}$ in the final electronic states with pronounced surface character (see surface state bands between 7.1\,eV and 7.3\,eV in Fig.\,S1 of Ref.\,[\onlinecite{SM}]).    
 Panel (b) shows the values of angle integrated $\gamma_{\bf k''}^{{\rm CFS}(2\hbar\omega_p)}$ on the energy scale, displaying a prominent peak at the energy $\sim E_F+2\hbar\omega_p$. Panel (c) shows the isoenergetic distribution of the values of $\gamma_{\bf k''}^{{\rm CFS}(2\hbar\omega_p)}$ across the (110) surface projected BZ obtained from averaging around $\epsilon_{\bf k''}\simeq E_F+2\hbar\omega_p$.\cite{averaged} The accumulations of significant contributions around the $\bar{\Gamma}$ and $\bar{\rm Y}$ points indicate a narrower phase space of final states than in the case of one plasmon-induced transitions displayed in Fig. \ref{hotelectron(110)}(c).

Taking into account only resonant processes the {\it joint} CFS rate $\gamma_{\bf k''}^{{\rm res}(2\hbar\omega_p)}$ of two plasmon-induced electron excitations is in the present situation obtained from (\ref{eq:2plasPE}) by retaining  only the on-the-energy shell components of intermediate state propagations contributing to this expression.\cite{NozieresI} This gives 
\bq
\gamma_{\bf k''}^{{\rm res}(2\hbar\omega_p)} \simeq \frac{\pi}{2}\rho_{f}(\epsilon_{\bf k''})\sum_{\bf k'} \gamma_{\bf k'',k'}^{(2\hbar\omega_p)}\gamma_{\bf k'}^{{\rm CFS}(\hbar\omega_p)}.
\label{eq:gamma2omega_p}
\eq
Here $\rho_{f}(\epsilon_{\bf k''})$ is the density of electron states around the final ${\bf k''}$-vector for which $\epsilon_{\bf k''}\simeq E_F+2\hbar\omega_p$, and $\gamma_{\bf k'}^{{\rm CFS}(\hbar\omega_p)}$ and $\gamma_{\bf k'',k'}^{(2\hbar\omega_p)}$ are defined in Eqs. (\ref{eq:GammaCFS}) and (\ref{eq:gamma2resolved}), respectively. The peculiar prefactor $\frac{\pi}{2}$ arises from the combination of different temporal boundary conditions governing the intermediate and final state electron propagation (i.e. over semi-infinite vs. infinite time interval, respectively). 

Using (\ref{eq:gamma2omega_p}) we can pinpoint the essential features of two plasmon-induced electron emission illustrated  diagramatically in Fig. \ref{plasmonyield}(b) and leading to non-Einsteinian electron yields.   
We first observe that the intensity bottleneck of resonant $2\hbar\omega_p$-electron emission described by the rate (\ref{eq:gamma2omega_p}) is the overlap of the phase space of intermediate resonant states $|{\bf k'}\rangle$ with the phase space of final states $|{\bf k''}\rangle$. This overlap strongly depends on the magnitude of plasmon wave vector cutoff. In the case of Ag(110) this is illustrated in Figs.  \ref{hotelectron(110)}(c) and \ref{hotelectron2(110)}(c) and favours the $2\hbar\omega_p$-emission in the directions of ${\bf k''}$ whose components $k_{x}''$ and $k_{y}''$ are restricted to the vicinity of $\bar{\Gamma}$ point. Here the trends noted in Fig. \ref{hotelectron2(110)} are further corroborated by the evaluation of consecutive two plasmon-induced resonant electron excitation rate (\ref{eq:gamma2omega_p}). The results are presented in Fig. \ref{res2wp_den}. Panel (a) shows the distribution of the values of $\gamma_{\bf k''}^{{\rm res}(2\hbar\omega_p)}$ within the (110)-surface projected BZ of Ag obtained from averaging around $\epsilon_{\bf k''}\simeq E_F+2\hbar\omega_p$.\cite{averaged} Panels (b) and (c) show histograms of the same data along the rectangular paths marked in the projected zone. All panels illustrate anisotropic accumulation of the largest values of $\gamma_{\bf k''}^{{\rm res}(2\hbar\omega_p)}$ around the $\bar{\Gamma}$ point, thereby favouring the majority of two plasmon-assisted photoemission intensity around the normal to the Ag(110) surface. The corresponding magnitudes of the rates characterizing the two plasmon-induced electron excitations indicate that these events  can take place within the duration of the ultrashort laser pulses. A good agreement between these theoretical predictions and  experimental findings presented in Refs. [\onlinecite{MarcelPRL,ACSPhotonics}] and Fig. \ref{AllAg} substantiates the mechanism of channeling of broad band optically induced electronic excitations into electron yield "monochromatized" to within the narrow range of plasmon dispersion.
   
\begin{figure}[tb] 
\rotatebox{0}{\epsfxsize=7.5cm \epsffile{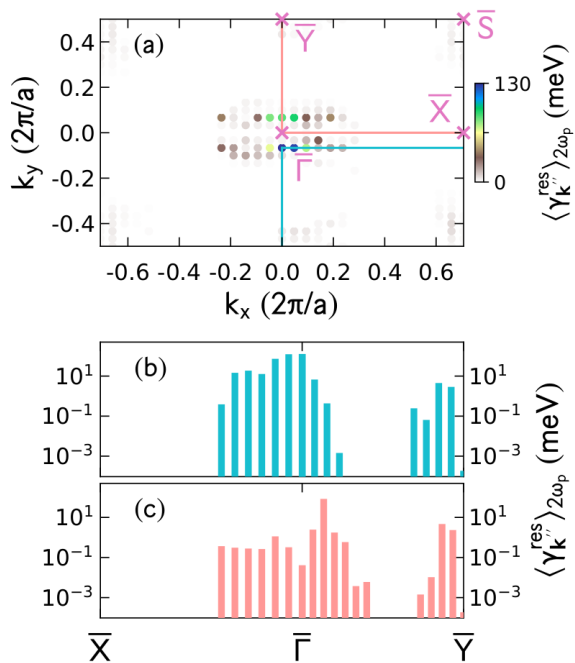}} 
\caption{(a): Distribution of the magnitudes of resonant two plasmon-induced electron excitation rates (\ref{eq:gamma2omega_p}) averaged around $\epsilon_{\bf k''}\simeq E_F+2\hbar\omega_p$ [\onlinecite{averaged}] and shown across the (110) surface-projected bulk BZ of Ag (see color bar for absolute values). (b) and (c): Histograms of the same data along the blue and red rectangular paths marked in (a).}
\label{res2wp_den}
\end{figure}

\section{Summary and outlook}
\label{sec:summary}

Following the above theoretical tour de force that establishes a mechanism for non-Einsteinin plasmonic photoemission, it is important to consider its practical implications. The decay of plasmons into hot electrons is an intensely invoked but hardly directly investigated theme in physics and chemistry.\cite{Narang2016,Marini2002,Atwater14,Nordlander15,Louie15,Halas15,deAbajo16,Atwater16,Khurgin} The experimental measurement of plasmonic photoemission\cite{MarcelPRL,ACSPhotonics}  and its theoretical interpretation presented herein imply that the excitation of the bulk plasmonic mode in the single crystal Ag samples may lead to transfer of the entire plasmon energy to selectively promote electrons from the Fermi level to $E_F+n\hbar\omega_p$, rather than to a distribution that is defined by the density of states and energy conservation, as is commonly believed.\cite{Narang2016,Marini2002,Atwater14,Nordlander15,Louie15,Halas15,deAbajo16,Atwater16,Khurgin}  There has been evidence that this selectivity occurs in alkali metals\cite{Gornyi,SmithSpicer,Smith,SmithFisher} and we have demonstrated that it also occurs in single crystalline Ag. If indeed the plasmonic response of metals can transfer energy entirely to hot electrons, rather than distributing it also among the holes, implies that plasmonic energy harvesting can be far more economical than commonly assumed. Therefore, it is important to consider whether the reported plasmonic photoemission is a peculiarity of the studied physical system or a common response of metals. The evidence so far suggests that it is indeed a plasmonic response of metals, based on what has been reported for alkali metals,\cite{Gornyi,SmithSpicer,Smith,SmithFisher} single crystalline Ag films,\cite{Horn} and Ag(111), (100), and (110) crystals.\cite{MarcelPRL,ACSPhotonics}  Our starting model, however, is not constrained by momentum conservation, so it does not preclude similar plasmonic decay into hot electrons for surface plasmon polaritons or bulk plasmons in nanostructured Ag. Indeed, a recent report on 2PP from size selected Ag clusters, where momentum conservation is relaxed,\cite{Douketis} confirms our claim of plasmonic photoemission from Ag.\cite{Shibuta} Moreover, our previous studies of two-photon photoemission from silver nanoclusters on TiO$_2$ and graphite surfaces, which were interpreted as originating from interface states,\cite{TanArgondizzo,TanLiu,TanDai} most likely include strong enhancement from plasmonic excitation of electrons from $E_F$. Thus, our experimental and theoretical evidence for plasmonic photoemission invites a broader reinterpretation of ultrafast electron dynamics in metals. To this end, we leave the open question whether an unresolved anomaly in hot electron dynamics in Cu\cite{HertelKnoesel,OgawaNagano,Pawlik,KnoeselHotzel,Weida} can be attributed to processes that could be interpreted as interband polarization having a similar role as plasmonic photoemission in Ag. In the case of Cu, the apparent hot electron lifetimes for photon excitation above the interband threshold have appeared to be anomalous, because they do not follow the Fermi liquid theory.\cite{OgawaNagano,Pawlik} Wolf and coworkers have attributed this anomaly to the d- to sp-band excitation,\cite{KnoeselHotzel} which enables the d-band holes to decay by Auger recombination to generate hot electrons having energies of $>2$ eV above $E_F$. While this can explain some results, there has never been clear evidence that there is a delayed rise of hot electrons reflecting the lifetimes of d-band holes.\cite{Weida} One can conceive that as the $d-$ to $sp$-band excitation channel becomes energetically accessible, the interband polarization of $d-$holes decays by a coherent process, analogous to plasmonic photoemission in Ag, by transferring the hole energy to electrons at $E_F$, rather than through an incoherent Auger process.

To summarize, we have studied the response of a system of coupled electrons and plasmons in a metal to perturbations exerted by a laser pulse on the electrons. We demonstrate that the wavefunction of the pulse-driven system incorporates the truncated plasmonic coherent states whose decay may generate complementary channels of plasmonically assisted electron emission from metal bands. This new paradigm enables us to elucidate  the origin and quantify the intensity of "non-Einsteinian" plasmon energy-locked electron yield from the Fermi surface observed in photoemission from Ag(110).\cite{ACSPhotonics} This reveals an as yet unexplored mode of "monochromatized" electron emission in optical energy conversion and harvesting\cite{Boriskina,Narang2016} from broad band radiation that so far has escaped the deserved scrutiny.  

This work is dedicated to the memory of Hiromu Ueba, scientist and friend, who pioneered the theory of 2PP from surfaces.\cite{UebaSS}

\acknowledgments

D. N. acknowledges the financial support from the Croatian Science Foundation (Grant no. UIP-2019-04-6869) and from the European Regional Development Fund for the ``Center of Excellence for Advanced Materials and Sensing Devices'' (Grant No. KK.01.1.1.01.0001).
A.L. and H.P. acknowledge the financial support from DOE-BES Division of Chemical Sciences, Geosciences, and Biosciences Grant No. DE-SC0002313. 
M.R. acknowledges funding from the Alexander von Humboldt Foundation.
Computational resources provided by the DIPC Computing Center are also acknowledged.


\appendix
\section{Nonperturbative evaluation of expression (\ref{eq:electronpart})}
\label{sec:ElectronPart}

To evaluate expression (\ref{eq:electronpart}) nonperturbatively we first introduce the interaction representation of the evolution operators with respect to the interaction $V^{e-pl}$ in (\ref{eq:Hsyst}) by

\barr
 U^{syst}(t,t')&=& U_{0}^{syst}(t,t')U_{I}^{syst}(t,t')\nonumber\\
&=& 
e^{-\frac{i}{\hbar}H_{0}^{syst}(t-t')}U_{I}^{syst}(t,t')
\label{eq:UI}
\earr
where

\bq
 U_{I}^{syst}(t,t')=1-\frac{i}{\hbar}\int_{t'}^{t}dt_{i} U_{I}(t,t_{i})V_{I}^{e-pl}(t_{i})
\label{eq:UIsyst}
\eq
with

\bq
 V_{I}^{e-pl}(t_{i})=e^{-\frac{i}{\hbar}H_{0}^{syst}(t_{i}-t')}V^{e-pl}e^{-\frac{i}{\hbar}H_{0}^{syst}(t_{i}-t')}.
\label{eq:V_I}
\eq
For later convenience we shall make use of the exponential representation of (\ref{eq:UIsyst}) in the form\cite{Gross}

\bq
 U_{I}^{syst}(t,t')=e^{-iG(t,t')}=\exp\left[-\frac{i}{\hbar}\sum_{l=1}^{\infty}g^{l}G_{l}(t,t')\right],
\label{eq:eG}
\eq
where $g$ is the coupling constant, $G(t,t')$ is a hermitian operator and the expansion terms $G_{l}(t,t')$ are obtained from a nested commutator expansion of the powers of interactions $V_{I}^{e-pl}(t')$ whose several lowest members are explicitly shown in Eqs. (183)-(186) of Ref. [\onlinecite{PhysRep}]. The leading term is

\barr
G_{1}(t,t')&=&\frac{1}{\hbar}\int_{t'}^{t}dt_{1}V_{I}^{e-pl}(t_{1})\nonumber\\
&=&
g\sum_{\bf \bar{k},k,q}V_{\bf \bar{k},k}^{\bf q}c_{\bf \bar{k}}^{\dag}c_{\bf k}a_{\bf q}^{\dag}\Delta_{\bf \bar{k},k}^{\bf q}(t,t')+h.c., \label{eq:G1}
\earr
with ${\bf k}$ and ${\bf \bar{k}}$ summations restricted such that $\epsilon_{\bf k}>\epsilon_{F}$ and   $\epsilon_{\bf \bar{k}}>\epsilon_{F}$. Here the phase integral $\Delta_{\bf \bar{k},k}^{\bf q}(t,t')$ defined in (\ref{eq:DeltaText}) 
arises from uncorrelated interactions induced by $U_{I}^{syst}(t,t')$. It turns into an energy conserving $\delta$-function in the adiabatic limit $(t'\rightarrow-\infty, t\rightarrow\infty)$. Analogous time-dependent quantities in all higher terms $G_{n>1}$ are much more complicated and give only the time correlated corrections to the elementary processes described by $G_{1}(t,t')$.\cite{PhysRep,Gross} In the following we shall effectively use the form 

\bq
U_{I}^{syst}(t,t')\simeq \exp\left[-i\left(gG_{1}(t,t')+g^{2}G_{2}^{diag}(t,t')\right)\right]
\label{eq:UG1G2}
\eq
because, as demonstrated in Sec. 4.3 of Ref. [\onlinecite{PhysRep}], this gives major contribution to (\ref{eq:electronpart}).

To proceed with evaluation of (\ref{eq:electronpart}) we consecutively substitute therein expressions (\ref{eq:UI}), (\ref{eq:UIsyst}) and (\ref{eq:eG}), and exploit the fact that $H_{0}^{syst}$ and $H_{0}^{pl}$ commute. This yields
%

{\small
\barr
&&
\langle 0|c_{\bf k_{1}}U^{syst}(t,t'')^{\dag}e^{-\frac{i}{\hbar}H_{0}^{pl}\tau}U^{syst}(t,t')c_{\bf k_{1}}^{\dag}|0\rangle=
\label{eq:UsystHUsyst}\\
&=&
\langle 0|c_{\bf k_{1}}U_{I}^{syst}(t,t'')^{\dag}e^{-\frac{i}{\hbar}H_{0}^{pl}\tau}U_{I}^{syst}(t,t')c_{\bf k_{1}}^{\dag}|0\rangle e^{-\frac{i}{\hbar}\epsilon_{\bf k_{1}}(t''-t')}\nonumber\\
\label{eq:UHU}\\
&=& 
\langle 0|c_{\bf k_{1}}(e^{-iG(t,t'')})^{\dag}e^{-\frac{i}{\hbar}H_{0}^{pl}\tau}e^{-iG(t,t')}c_{\bf k_{1}}^{\dag}|0\rangle e^{-\frac{i}{\hbar}\epsilon_{\bf k_{1}}(t''-t')}\nonumber\\
\label{eq:GHG}\\
&=&
\langle 0|c_{\bf k_{1}}\exp[-i (e^{-iG(t,t'')})^{\dag}H_{0}^{pl}e^{-iG(t,t')}\tau]c_{\bf k_{1}}^{\dag}|0\rangle e^{-\frac{i}{\hbar}\epsilon_{\bf k_{1}}(t''-t')}.\nonumber\\
\label{eq:<el>}
\earr
}
%
Thereby we have reduced expression (\ref{eq:UsystHUsyst}) to the form of excited state average of a generalized exponential operator\cite{alternative} (\ref{eq:<el>}) and this can be evaluated using cumulant expansion.\cite{PhysRep,Kubo} To apply the latter we must find the explicit form of the transformed operator $H_{0}^{pl}$ in the exponent in (\ref{eq:<el>}), viz. of the product

\bq
e^{iG(t,t'')}H_{0}^{pl}e^{-iG(t,t')}=U_{I}^{syst}(t,t'')^{\dag}H_{0}^{pl}U_{I}^{syst}(t,t').
\label{eq:calHpl}
\eq
For $t'\neq t''$ this operator is {\it not} symmetric in temporal variables and therefore not amenable to the simple operator algebra based on the expansion 

\bq
e^{iB}A e^{-iB}=A+\sum_{n=1}\frac{i^{n}}{n!}B^{n}[A] 
\label{eq:commutators}
\eq
which is routinely employed in manipulations with evolution operators. Here $B^{n}[A]$ denotes the $n$-th order repeated commutator of operator $B$ with operator $A$.  However, we observe that for ultrashort pulses the maximum contribution to the overall expression (\ref{eq:Nplkk}) arises during the overlap of the pulse amplitudes ${\cal E}(\omega_x,t',\sigma_{x})$ and  ${\cal E}(\omega_x,t",\sigma_{x})$, say at $t'\sim t''\sim t_{x}$. To exploit this we introduce new temporal coordinates, the pumping mean time $T$ and the relative or mismatch time $\bar{t}$ by 

\bq
T=\frac{t''+t'}{2}, \hspace{1cm} \bar{t}=t''-t'.
\label{eq:times}
\eq
This is analogous to Keldysh's introduction of "slow" ($T$) and "fast" ($\bar{t}$) variables, whereupon the "fast" one is integrated out in his derivation of the kinetic equation.\cite{Keldysh} 
It is seen that the mean or central time $T$ is strongly restricted to the interval $\sigma_{x}$ around $t_{x}$ and  the relative time $\bar{t}$ to the values around zero, i.e. $\bar{t}<\sigma_{x}$. In terms of these coordinates we obtain for the RHS of (\ref{eq:calHpl}) a partially symmetric expression

\begin{widetext}
\barr
U_{I}^{syst}(t,t'')^{\dag}H_{0}^{pl}U_{I}^{syst}(t,t')&\rightarrow& U_{I}^{syst}(T+\bar{t}/2,T) U_{I}^{syst}(T,t)H_{0}^{pl}U_{I}^{syst}(t,T)U_{I}(T,T-\bar{t}/2)\nonumber\\
&\sim& 
U_{I}^{syst}(t,T)^{\dag}H_{0}^{pl}U_{I}^{syst}(t,T) + {\cal O}({\bar t}), 
\label{eq:calHdiag}
\earr
\end{widetext}
where the "temporally diagonal" component in the last line on the RHS of 
 (\ref{eq:calHdiag}) picks the maximum contribution to (\ref{eq:Nplkk}) and becomes exact in the limit of extreme ultrashort pulses, i.e.  ${\cal E}(\omega_x,t,\sigma_{x})\propto \delta(t-t_{x})$. Hence, in the following we shall retain only this leading term. The latter is now transformed using (\ref{eq:commutators}) to give 

\bq
U_{I}^{syst}(t,T)^{\dag}H_{0}^{pl}U_{I}^{syst}(t,T)=H_{0}^{pl}+{\cal W}(t,T)
\label{eq:calH}
\eq
which is now substituted in (\ref{eq:<el>}). 
Thereby expression (\ref{eq:electronpart}) is reduced to the excited state average of a single exponential operator, viz. 

\bq
\langle 0|c_{\bf k_{1}}\exp\left[-\frac{i}{\hbar}\left(H_{0}^{pl}+{\cal W}(t,T)\right)\tau\right]c_{\bf k_{1}}^{\dag}|0\rangle
\label{eq:<eW>}
\eq
where according to (\ref{eq:commutators}) the "interaction" ${\cal W}(t,T)$ is expressible as yet another nested commutator series\cite{PhysRep,Gross}    

\bq
{\cal W}(t,T)=\sum_{n=1}^{\infty}\frac{i^n}{n!}G^{n}(t,T)[H_{0}^{pl}]. 
\label{eq:calW}
\eq
Due to the very complex structure of the various $G_{n}$'s constituting (\ref{eq:eG}) the series (\ref{eq:calW}) is likewise complex and hence we shall again retain only the terms of lowest order in the electron-plasmon coupling constant $g$. Thus we obtain for the leading term

\bq
{\cal W}_{1}(t,T)=-ig\sum_{\bf \bar{k},k,q}\hbar\omega_{\bf q}V_{\bf \bar{k},k}^{\bf q}c_{\bf \bar{k}}^{\dag}c_{\bf k}a_{\bf q}^{\dag}\Delta_{\bf \bar{k},k}^{\bf q}(t,T) + h.c.
\label{eq:calW1}
\eq
The final stage in bringing the RHS of (\ref{eq:<eW>}) to a menagable form follows from the observation that this expression has a formal appearance of the excited state average of a Heisenberg evolution operator in the $\tau$-space (recall that $\tau$ is not the genuine evolution time). This means that we can express it in the interaction picture in the same $\tau$ space, viz. in the form $e^{-\frac{i}{\hbar}H_{0}^{pl}\tau}U_{I}^{\cal W}(\tau)$ and thereby reduce the RHS of (\ref{eq:<eW>}) to the expression amenable to cumulant treatment\cite{Kubo,GW+C} in the $\tau$-space. This procedure is described in detail in Sec. 4.3 of Ref. [\onlinecite{PhysRep}] and here we only quote the final result 

\begin{widetext}
\bq
 \langle 0|c_{\bf k_{1}}\exp\left[-\frac{i}{\hbar}\left(H_{0}^{pl}+{\cal W}(t,T)\right)\tau\right]c_{\bf k_{1}}^{\dag}|0\rangle\rightarrow 
\exp\left[-g^{2}\sum_{\bf \bar{k}_1, q}\left|V_{\bf \bar{k}_1,k_{1}}^{\bf q}\Delta_{\bf \bar{k}_1,k_{1}}^{\bf q}(t,(t'+t'')/2)\right|^{2}(1-n_{\bf \bar{k}_1,})(1-e^{-i\omega_{\bf q}\tau})\right]
\label{eq:<eWfinal>}
\eq
\end{widetext}
where on the RHS we have restored the original time evolution variables  $t'$ and $t''$. This representation holds within the validity of second order cumulant expansion of the electron propagator.\cite{4cumul}



\begin{thebibliography}{11}
\bibitem{Glauber} These states are completely analogous to photonic coherent states excited by electronic motion (current) as described by Eqs. (9.16)-(9.23) in Ref. [\onlinecite{GlauberPR}].
\bibitem{Sakaue} An illustrative discussion of the importance of establishing a hierarchy of the various characteristic times of the system explored by short laser pulses was presented in Ref. [\onlinecite{SakaueJP}].
\bibitem{RefCaroli} The significance of analogous processes assisted by phonons instead of plasmons was pointed out in Ref. [\onlinecite{Caroli}], cf. Fig. 9(b) therein.
\bibitem{NozieresI} The situation in which the on-the-energy-shell imaginary parts of intermediate state electron propagators dominantly determine the optical properties of metals was discussed in Sec. III of Ref. [\onlinecite{Nozieres69I}].
\bibitem{alternative} Reduction to a single exponential operator may also be started from expression (\ref{eq:GHG}) by a repeated application of the Baker-Campbell-Hausdorff formula
 $ e^{X}e^{Y}=e^{Z}$ where $ Z=X+Y+\frac{1}{2}[X,Y]+\frac{1}{12}[X,[X,Y]]+\cdots $, but  this yields  slower convergent results regarding the functional dependence on $\tau$ than the procedure followed in the text.
\end{thebibliography}
\end{document}